\documentclass[table, journal]{IEEEtran}

\def\BibTeX{{\rm B\kern-.05em{\sc i\kern-.025em b}\kern-.08em T\kern-.1667em\lower.7ex\hbox{E}\kern-.125emX}}
\usepackage[numbers]{natbib}

\usepackage[switch]{lineno}
\usepackage{stfloats}
\usepackage{graphicx}

\usepackage{indentfirst}
\usepackage{array,multirow}
\usepackage{fancyhdr}
\usepackage{float}
\usepackage{booktabs}
\usepackage{here}
\usepackage{comment}
\usepackage{siunitx}
\usepackage{algorithmicx,algorithm,algpseudocode}
\usepackage{amsmath,amssymb,amsfonts,bm,gensymb}
\usepackage{fontawesome5}
\usepackage{caption}
\usepackage{setspace}
\usepackage{xcolor}
\usepackage{cite}
 
\usepackage{svg}
\usepackage{media9}
\newcolumntype{R}{>{\centering\arraybackslash}m{0.8cm}}
\usepackage[normalem]{ulem}
\useunder{\uline}{\ul}{}
\usepackage{url,hyperref,microtype}

\usepackage{hhline} 
\usepackage{highlight}
\renewcommand{\opacity}{90}

\newcommand{\ctext}[1]{\raise0.2ex\hbox{\textcircled{\scriptsize{#1}}}}

\newcommand{\etal}{et~al.\ }
\newcommand{\eg}{e.\,g.,\ }
\newcommand{\ie}{i.\,e.,\ }

\begin{document}

\title{
Data-driven Causal Discovery for Pedestrians-Autonomous Personal Mobility Vehicle Interactions with eHMIs: From Psychological States to Walking Behaviors
}

\author{
Hailong~Liu$^{1,*}$,~\IEEEmembership{Senior Member, IEEE}, Yang~Li$^{2}$, \\
Toshihiro~Hiraoka$^{3}$,~\IEEEmembership{Member, IEEE} and  Takahiro~Wada$^{1}$,~\IEEEmembership{Member, IEEE}
\vspace{-6mm}
\thanks{
$^{1}$Hailong~Liu and Takahiro~Wada are with the Graduate School of Science and Technology, Nara Institute of Science and Technology, 8916-5 Takayama-cho, Ikoma, Nara 630-0192, Japan.}

\thanks{
$^{2}$Yang~Li is with Institute of Human and Industrial Engineering, Karlsruhe Institute of Technology, Engler-Bunte-Ring 4, Karlsruhe, 76133, Germany.}

\thanks{
$^{3}$Toshihiro~Hiraoka is with Japan Automobile Research Institute, 1-1-30, Shibadaimon, Minato-ku, Tokyo 105-0012, Japan.}

\thanks{*CONTACT Hailong Liu. \faIcon[regular]{envelope}~:~{\tt\small liu.hailong@is.naist.jp}}
}
\maketitle

\begin{abstract}
Autonomous personal mobility vehicle (APMV) is an innovative small autonomous transportation device designed for individual use in mixed-traffic environments, such as shared spaces and indoor environments. 
To enhance the interaction experience between pedestrians and APMVs and to prevent potential risks, it is crucial to investigate pedestrians' walking behaviors when interacting with APMVs and to understand the psychological processes underlying these behaviors.
This study aims to investigate the causal relations between subjective evaluations of pedestrians and their walking behaviors during interactions with an APMV equipped with an external human-machine interface (eHMI).
An experiment of pedestrian-APMV interaction was conducted with 42 pedestrian participants, in which various eHMIs on the APMV were designed to induce participants to experience different levels of subjective evaluations and generate the corresponding walking behaviors.
Based on the hypothesized model of the pedestrian's cognition-decision-behavior process, the results of causal discovery align with the previously proposed model.
Furthermore, this study further analyzes the direct and total causal effects of each factor and investigates the causal processes affecting several important factors in the field of human-vehicle interaction, such as situation awareness, trust in vehicle, risk perception, hesitation in decision making, and walking behaviors.

\end{abstract}


\section{INTRODUCTION}

\begin{figure}[t]
    \centering
    \includegraphics[width=1\linewidth,trim=14 0 2 0,clip]{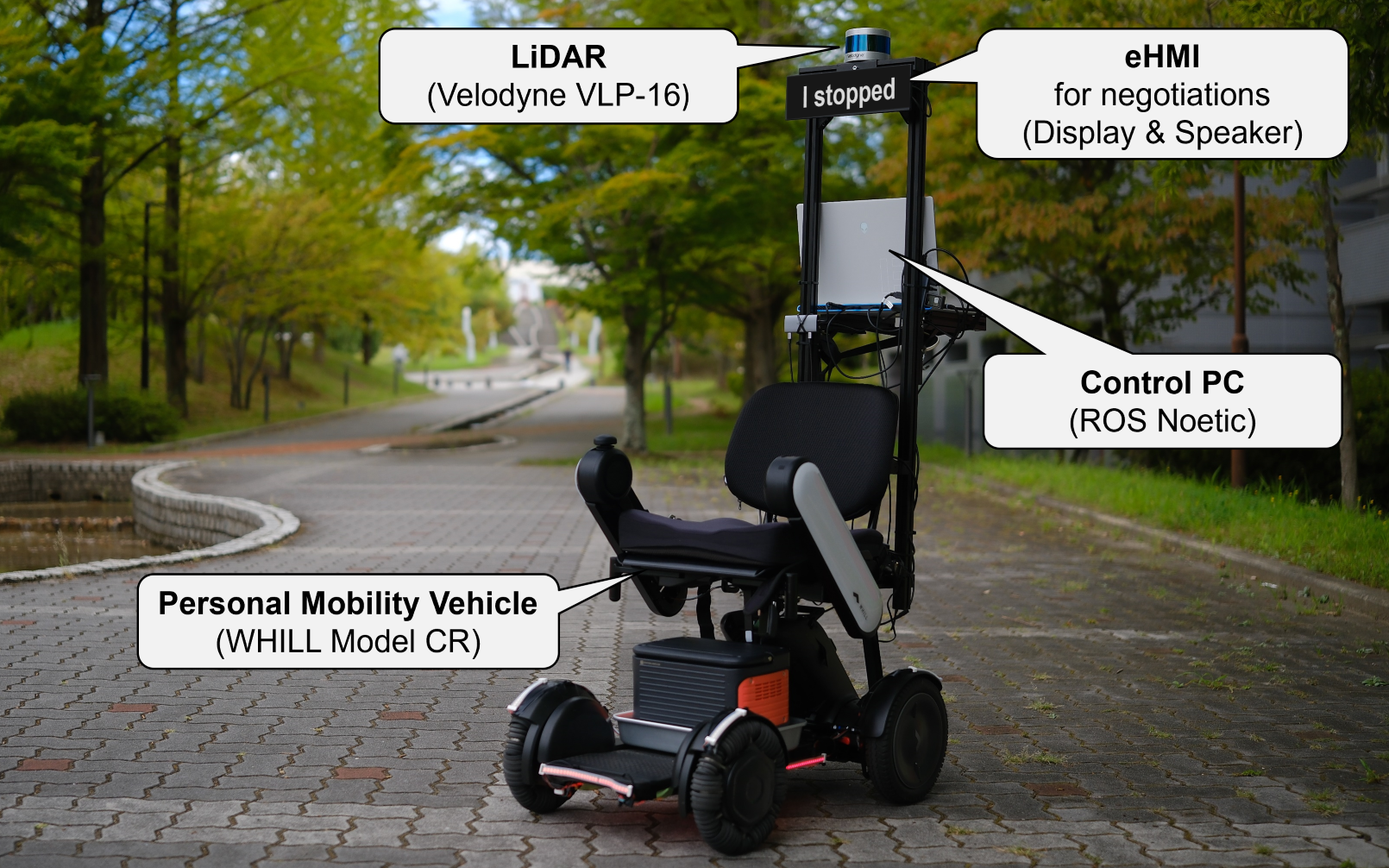}
    \caption{The APMV with an eHMI can convey driving intentions to pedestrians through voice and visual cues.}
    \label{fig:APMV}
\end{figure}

\IEEEPARstart{A}{utonomous} personal mobility vehicles (APMVs) are innovative small autonomous transportation devices designed to provide convenient, safe, and efficient mobility for individuals~\citep{liu2022implicit, liu2024_APMV_eHMI} (see Fig.~\ref{fig:APMV}).
APMVs typically feature autonomous driving levels ranging from SAE levels 3 to 5~\citep{SAE_j3016b_2018}. 
While many well-known APMVs are currently developed based on electric wheelchairs~\footnote{\textit{WHILL Autonomous} developed by WHILL Inc.: \url{https://youtu.be/vJWhwNnUPRs}} or semi-open small vehicles~\footnote{\textit{RakuRo} developed by ZMP Inc.: \url{https://youtu.be/lWhbJ0rBwjM}}, APMVs are not exclusively designed for the elderly or people with disabilities.
Indeed, APMVs aim to address the ``last one mile'' problem and enhance short-distance mobility for a broader range of users~\citep{liu2024_APMV_eHMI}.
APMVs usually operate at low speed, especially in environments shared with pedestrians or other non-motorized vehicles.
For example, APMVs can serve as last-mile transportation solutions in urban areas~\citep{morales2017social}, connecting public transit stops to final destinations within city centers, commercial zones, or residential neighborhoods.
They are also ideal for short-distance commutes within university campuses~\citep{liu2022implicit,watanabe2015communicating, andersen2016autonomous}, parks, or shopping centers~\citep{morales2018personal,isono2022autonomous}.
As a result, APMVs will inevitably and frequently interact with other road users, such as pedestrians~\citep{liu2024_APMV_eHMI,yoshitake2023pedestrian,liu2024subjective}.

 \begin{figure*}[h!t]
    \centering
    \includegraphics[width=1\linewidth]{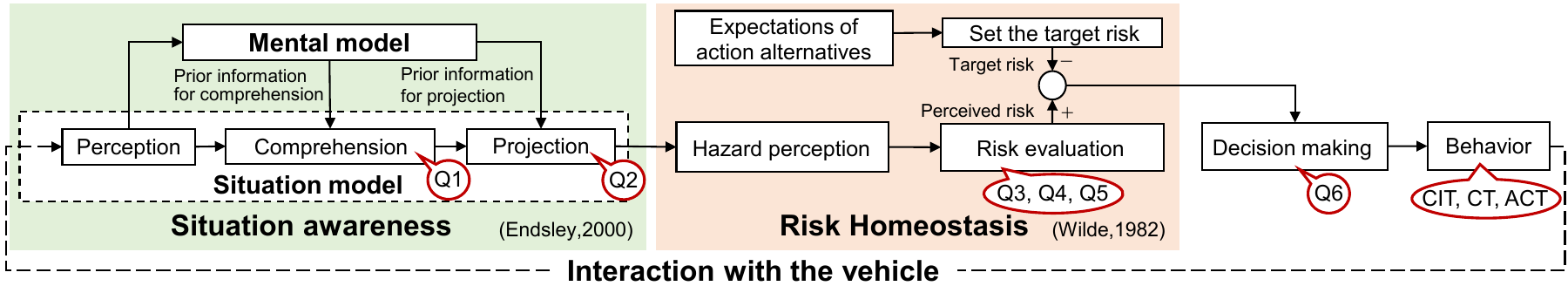}
    \caption{A hypothesized model of pedestrian's cognition-decision-behavior process in human-vehicle interactions was proposed in~\citep{liu2020_what_timeing,liu2022implicit}. The Q1 to Q6 are subjective evaluation questionnaires using in the experiments; and the crossing initiation time (CIT), crossing time (CT) and after crossing time (ACT) are the time spent on the three stages of crossing the road (see section~\ref{sec:Measurements}).}
    \label{fig:model}
 \end{figure*}


\subsection{Communication Issues in Pedestrians-APMV Interactions}

In some studies, potential risks have been identified in the interaction between APMVs and pedestrians.
For example, \citet{liu2022implicit,liu2020_what_timeing} reported that pedestrians tended to perceive danger because they were confused in understanding the driving intentions of the APMV during the interaction.
\citet{yamin2023did} also emphasizes that a lack of effective communication between the APMV and pedestrians can lead to potential collision risks for pedestrians.

To address this challenge, the integration of an External Human-Machine Interface (eHMI) into the APMV has been identified as an effective approach to convey driving intentions to pedestrians~\citep{liu2024_APMV_eHMI,watanabe2015communicating}.

\subsection{Related Studies}
\subsubsection{Human-AV (APMV) interaction with eHMI}

At present, a significant amount of studies on eHMI focus on interactions between pedestrians and autonomous vehicles (AVs)~\citep{chang2017eyes, Clercq2019eHMI, burns2019pedestrian, liu2021importance,loew2022go, lee2022learning, dey2021communicating,song2024road,liu2025pre}. 
Only a limited number of studies have focused on interactions between small types of smart mobility vehicles, such as APMVs, with pedestrians in shared spaces using eHMIs~\citep{liu2024_APMV_eHMI,zhang2024shared,liehr2024you}.

These studies have drawn conclusions about the influence of eHMI usage on pedestrians' subjective evaluations and behaviors by comparing different experimental conditions, such as various eHMIs and driving behavior scenarios.
For example, information from eHMIs has been shown to help pedestrians understand the vehicle's intention~\citep{liu2021importance,dey2021communicating,song2024road,liu2025pre,liehr2024you}, enhance their sense of safety~\citep{Clercq2019eHMI,song2024road,liu2025pre,faas2020external}, foster trust in the AV~\citep{dey2021communicating,liu2025pre,faas2020external}, support decision-making~\citep{chang2017eyes,burns2019pedestrian,liu2025pre}, improve crossing initiation time~(CIT)~\citep{loew2022go,lee2022learning}, and guide walking behaviors~\citep{liu2025pre}.
In contrast, some studies~\citep{hollander2019overtrust,kaleefathullah2020external} showed that eHMIs might lead to over-trust from pedestrians toward AVs, especially when pedestrians fail to correctly interpret the AVs' intentions from the eHMI, or over-interpret the eHMI information.
This over-trust could induce risky behaviors of pedestrians during interactions with AVs, thereby increasing the likelihood of accidents.

Although the aforementioned studies have explored how the use of eHMIs in interactions between pedestrians and AVs or APVMs could improve pedestrians' subjective experiences and behavioral responses, and might induce inappropriate levels of trust, 
their findings were primarily based on comparative and correlational analyses, thus the underlying mechanisms behind these effects remain unclear.
Specifically, the causal relations, including causal paths, effects, and processes among pedestrians' psychological states and their walking behaviors during interactions with AVs or APVMs, remain insufficiently validated or explored.
Furthermore, the examination of interactions involving pedestrians and APMVs is even less prevalent in existing research.

\subsubsection{Causal relation Analysis}

Structural Equation Modeling (SEM) is primarily used to test and estimate parameters of a predefined causal structure based on a theoretical model~\citep{tomarken2005structural}. 
SEMs are commonly used to analyze and validate causal relations between multiple variables in studies of human-robot interactions~\citep{kim2020factors,ruiz2023human,chen2024exploring} and human-vehicle interactions~\citep{nordhoff2021interrelationships,li2024effect}.

Unlike conventional SEMs, causal discovery methods aim to automatically learn the causal structure from observational data~\citep{spirtes2016causal,glymour2019review}.
These methods include constraint-based methods such as the Peter-Clark (PC) algorithm~\citep{spirtes1991algorithm}, score-based methods such as Greedy Equivalence Search (GES)~\citep{chickering2002optimal}, and SEM-based approaches such as the Linear Non-Gaussian Acyclic Model (LiNGAM)~\citep{shimizu2006LinGAM}.
Currently, causal discovery is widely used across various research fields, such as 
psychology~\citep{mojtabai2024problematic}, economics~\citep{bruns2021estimating}, neuroscience~\citep{tu2019neuropathic}, as well as astronomy~\citep{jin2025causal}.

In the field of pedestrian–APMV interaction, a preliminary work of this study has been published in \citep{liu2024causal}, which presents an initial causal discovery modeling pedestrians' psychological states to their walking behaviors.
In this preliminary work, we invited 18 participants as pedestrians to interact with the APMV. 
Through causal discovery, we found some direct causal relations between some factors of subjective evaluations and walking behaviors.
However, due to the small number of participants, some direct causal relations of subjective factors, such as causal effects related to trust and perceived danger, are not statistically stable and have low reproducibility via bootstrap.

\subsection{Purpose}

Based on a large-scale subject experiment with 42 participants, this study aims to discover and analyze the causal relations between subjective evaluations of pedestrians and their walking behaviors during interactions with an APMV via a data-driven approach, including scenarios where the APMV communicates with pedestrians using eHMIs as well as scenarios without eHMI communication.
Furthermore, this study aims to investigate the previously proposed hypothesized model (see Fig.~\ref{fig:model}) based on data-driven causal discovery results.

\subsection{Contributions}

The novelty of this study is that it employs a data-driven causal discovery approach to uncover a causal process from pedestrians' subjective evaluations to their walking behaviors during interactions with APMVs.
In addition, we further analyze the total causal effects of each factor and investigate the causal processes affecting several important factors, such as trust in the vehicle, in the field of human-vehicle interaction.

The contributions of this study are as follows:
\begin{itemize}
\item[1)] A data-driven causal discovery approach was used to investigate a hypothesized model of pedestrians' cognition-decision-behavior process in human-vehicle interaction (see Fig.~\ref{fig:model}).
\item[2)] A direct causal effect of situational awareness on trust in APMV was identified, highlighting the need to calibrate trust by providing information that improves situational awareness and helps pedestrians form an accurate mental model. 
\item[3)] This study found that when the APMV conveyed its driving intentions to pedestrians through eHMI during or before changes in driving behavior, it enhanced subjective evaluations of pedestrians and made their walking behaviors more efficient during interactions.
\end{itemize}

\section{CAUSAL DISCOVERY}

\subsection{Hypothesized Model of Pedestrian's Cognition-Decision-Behavior Process}

To discover the causal relations between the subjective evaluations of pedestrians and their walking behaviors during interactions with APMV, this study adopts the conceptual model of the pedestrian cognition-decision-behavior process proposed by~\citep{liu2020_what_timeing,liu2022implicit}
as the hypothesized model.

As shown in Fig.~\ref{fig:model}, the hypothesized model consists of four fundamental sections, namely \ie awareness of the situation~\citep{endsley1995toward}, risk homeostasis~\citep{wilde1982theory}, decision making and behavior generation.
The situation awareness section describes the cognitive processes of pedestrians, including perception, comprehension, and projection, which together form their overall awareness of the surrounding environment~\citep{endsley1995toward}.
This process is supported by the mental model, which provides the necessary prior knowledge for situation awareness~\citep{endsley2000}.
Afterwards, in the risk homeostasis section, pedestrians assess hazards based on predictions and evaluate subjective risks (\eg sense of danger, sense of relief) in the current situation, taking into account their personal level of risk acceptance (\ie target risk).
Furthermore, \citet{liu2021importance} suggested that pedestrians' trust in the APMV interacts with the target risk, influencing their subjective risk evaluations.
Following this, pedestrians compare their subjective risk with the acceptable risk level (\ie the target risk) to inform  their decision-making. 
Once this decision-making process is complete, the body executes specific walking behaviors.
As pedestrians engage in these walking behaviors, they interact with objects, \eg APMV, in their surrounding environment, thereby perpetuating the aforementioned process in a continuous loop.

Importantly, it should be noted that pedestrians' behaviors dynamically influence their future situation awareness through environmental interactions, as depicted by the dashed line in Fig.~\ref{fig:model}.
This indirect effect or indirect causal relation lies outside of this study's scope. 
Therefore, we assume that the direct causal process from pedestrians' perception in situational awareness to their behaviors is unidirectional and non-cyclical. 
This relation can be adequately represented by a directed acyclic graph (DAG).

\subsection{Causal Discovery via Direct Linear Non--Gaussian Acyclic Model (DirectLiNGAM)}

Causal discovery aims to uncover and understand the relations between variables from observed data, estimating which variables have direct or indirect causal effects on others.
In which, \citet{shimizu2006LinGAM} proposed the LiNGAM for estimating DAG-based SEM by using non-Gaussianity of the data.
Thus, LiNGAM is selected as the causal discovery method in this study because it is applicable to DAGs, which align with the structure of the hypothesized model (see Fig.~\ref{fig:model}).

The LiNGAM presents a DAG as
\begin{eqnarray}
\underbrace{
 \begin{bmatrix}
x_1\\
\vdots\\
\vdots\\
x_n
\end{bmatrix}
}_{\bm{x}}
&=&
\underbrace{
\begin{bmatrix}
0&\cdots&\cdots&0\\
a_{2,1}&\ddots&&\vdots\\
\vdots&\ddots&\ddots&\vdots\\
a_{n,1}&\cdots&a_{n,n-1}&0\\
\end{bmatrix}
}_{\bm{A}}
\underbrace{
\begin{bmatrix}
x_1\\
\vdots\\
\vdots\\
x_n
\end{bmatrix}
}_{\bm{x}}
+
\underbrace{
\begin{bmatrix}
e_1\\
\vdots\\
\vdots\\
e_n
\end{bmatrix}
}_{\bm{e}}
\label{eq:SEM}
\end{eqnarray}
in which $\bm{x}\in\mathbb{R}^n$ denotes the observed variables, with its subscript specifying the causal order. 
The $n$ is the number of observed variables.
The strictly lower triangular matrix $\bm{A}\in\mathbb{R}^{n\times n}$ indicates an adjacency matrix and the $\bm{e}$ indicates independent error variables. 
In matrix $\bm{A}$, $a_{i,j}$ represents the strength of the direct causal relation from $x_i$ to $x_j$, thereby illustrating the causal path and causal effect between these variables.

Generally, the objective of causal discovery using SEM is to estimate an optimal matrix $\bm{A}$ that characterizes the data generative process, \ie the causal relations among the observed data.
Since LiNGAM assumed that the causal process can be represented by a DAG, it permutes the matrix $\bm{A}$ to a strictly lower triangular matrix by simultaneous equal row and column permutations.
The lower triangular matrix $\bm{A}$ could be estimated using the independent component analysis~(ICA)~\citep{shimizu2006LinGAM}.
However, most iterative method-based ICA algorithms, such as FastICA, may depend on the initial parameter states, making it challenging to guarantee that LiNGAM will converge to the correct solution in a finite number of steps~\citep{shimizu2011directlingam}.

\begin{algorithm}[h!t]
\setstretch{0.9}
\caption{DirectLiNGAM (\citet{shimizu2011directlingam})}
\label{alg:DirectLiNGAM}
 \textbf{Input:} $\bm{x} = (x_1, x_2, \dots, x_n)^T\in\mathbb{R}^{n}$ where $n =\text{dim}(\bm{x})\in\mathbb{Z}^{+}$.\\
\textbf{Output:} Adjacency matrix $\bm{A}$ in Eq.~\ref{eq:SEM}.

\begin{algorithmic}[1]
\vspace{1mm}
\Statex \hspace{-5mm} \textbf{Stage 1: Determining the causal orders}
\vspace{1mm}
\While {$\text{dim}(\bm{x})>0$} 
    \State Perform least squares regressions of $x_i$ on $x_j$ and compute the residual $r_{ij}$:\vspace{-2mm}
    \[
    r_{ij} = x_i - \frac{\text{cov}(x_i, x_j)}{\text{var}(x_j)} x_j
    \]\vspace{-2mm}
    \State  Find a variable $x_m \in \mathrm{x}$ that is most independent of its residuals via mutual information (MI):\vspace{-2mm}
    \[
    x_m = \underset{j}{\text{arg min}} \sum_{i \neq j} MI(x_j, r_{ij}) 
    \]\vspace{-2mm}

   \State  Removing the influence of the exogenous variable $x_m$ on other variables. The other variables are update by: \vspace{-2mm}
\[
    x_i \gets r_{im}
    \]\vspace{-5mm}
    \State  Record the order of exogenous variable $x_m$ and remove it from $\bm{x}$, then $\bm{x}\in\mathbb{R}^{n\gets n-1}$.
    \EndWhile
\vspace{1mm}
\Statex \hspace{-5mm} \textbf{Stage 2: Optimizing the adjacency matrix $\bm{A}$}
\vspace{1mm}
\State  Sorting variables in $\bm{x}_{sort}$ based on their causal orders.
\State Optimizing the adjacency matrix $\bm{A}$ with input $\bm{x}_{sort}$ via a linear regression (e.g., Lasso) for Eq.~\ref{eq:SEM}.

\end{algorithmic}
\end{algorithm}

To address this issue, Shimizu \etal proposed DirectLiNGAM which can directly extract causal structures from observed data~\citep{shimizu2011directlingam}.
As shown in Algorithm~\ref{alg:DirectLiNGAM},
DirectLiNGAM can be divided into two stages: 1) determining the causal order between variables, 2) optimizing the adjacency matrix $\bm{A}$.

In the first stage, DirectLiNGAM identifies an exogenous variable that is most independent of the other variables, by minimizing the mutual information between the variables and their residuals (steps 2 and 3).
Once the exogenous variable is identified, the remaining variables are updated using the residuals between them and the exogenous variables, effectively removing the exogenous variables' influence (step 4).
Next, the identified exogenous variable is recorded and excluded from $\bm{x}$ (step 5). 
Then, DirectLiNGAM continues with the remaining variables to identify additional exogenous variables until $\text{dim}(\bm{x})=0$ (steps 1 to 6).
By iteratively repeating these steps, the causal orders among variables can be determined.

In the second stage,
once the causal orders of all variables are identified,
DirectLiNGAM will rearrange the order of variables in the vector$\bm{x}$ (step 7).
After that, DirectLiNGAM uses least squares regression to calculate the adjacency matrix $\bm{A}$, ensuring that it maintains a strictly lower triangular form in accordance with the determined causal order (step 8).
For more details on the DirectLiNGAM algorithm, please refer to \citep{shimizu2011directlingam}.

In this paper, as shown in Fig.~\ref{fig:model}, we assume that the data generation process, \ie causal process, from subjective evaluations to walking behaviors of pedestrians is unidirectional and non-cyclical, indicating that it can be represented by a DAG.
Consequently, DirectLiNGAM is used in this study.

\section{EXPERIMENT}

A within-subjects experiment was conducted to measure pedestrians' psychological states and walking behaviors during interactions with an APMV, and to discover the causal relation among these measured variables.
To support robust causal discovery, it is essential that the data set includes a wide range of psychological states, ranging from positive (\eg driving intention is easy to understand; and feeling very safe) to negative (\eg driving intention is difficult to understand; feeling very unsafe), along with the corresponding behavioral responses.
Therefore, four different eHMI conditions were designed to induce varying levels of situation awareness in pedestrians during the interaction.

Moreover, this experiment simulated a scenario in which the APMV autonomously drives to pick up passengers or returns to standby after dropping them off, i.e., operates autonomously without a passenger, in order to induce pedestrians' risk perception.
This is because pedestrians' trust in the passenger on the APMV could bias their risk perceptions, as discussed in \citet{liu2022implicit}.
For example, pedestrians may assume that the passenger can take over control at any time, leading them to perceive the risk as low regardless of how dangerous the APMV's driving behaviors might actually be.
The specific experimental design is detailed in the following subsection.

This experiment was carried out with the approval of the Research Ethics Committee of Nara Institute of Science and Technology, Japan (No.~2022-I-55-1).

\subsection{Participants}
An a priori power analysis using \textit{G*Power}~\citep{GPower} (version=3.1.9.7, effect size $f=0.25$, $\alpha=0.05$, $power=0.85$) estimated the minimum sample size for repeated measures ANOVAs under the four eHMI conditions was 26.
In this study, we invited 42 participants (self-reported genders: male 34, female: 8) with ages from 22 to 38 years (Avg.: 26.69 years, Std.: 4.36 years) from multiple countries to participate in the experiment.
There are 13 participants from Japan, 8 from Indonesia, 5 from China, 3 each from Malaysia and Bangladesh, 2 each from Pakistan, Philippines and Vietnam, and 1 participant each from Lebanon, Colombia, Sudan, and Nigeria.
None of the participants had previous experience using or interacting with APMVs.
The experiment lasted approximately one hour and each participant was compensated with 1,000 Japanese Yen for their participation.

\subsection{APMV and Its eHMI Device}

As shown in Fig.~\ref{fig:APMV}, a WHILL Model CR robotic wheelchair equipped with an autonomous driving system was used as the APMV. 
This APMV is equipped with a Velodyne VLP-16 LiDAR and a control PC for autonomous driving on a pre-designed route.
A display with a speaker has been installed on top of the APMV to function as the eHMI. 
This eHMI communicates driving intentions by showing relevant text on the display and vocalizing the messages through the speaker.

\subsection{Experimental Site}

\begin{figure*}[t]
  \centering
   \includegraphics[width=0.85\linewidth]{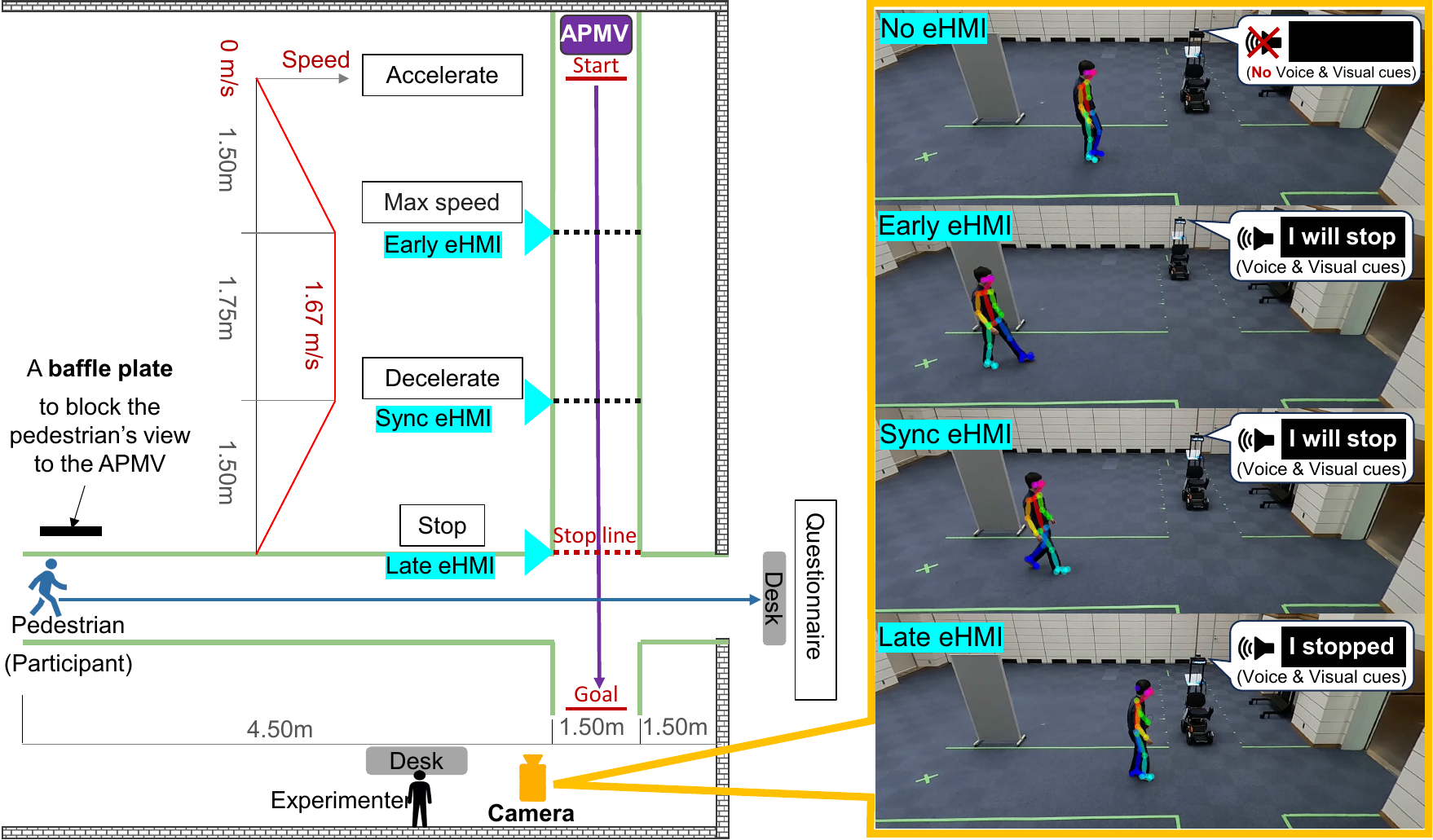}
  \caption{Experimental settings including experimental site, driving behavior profile of the APMV and eHMI conditions}
  \label{fig:scenarios}
\end{figure*}

An indoor area measuring 10~m$\times$10~m was arranged to simulate a street crossing scene, as shown in Fig.~\ref{fig:scenarios}.
The pedestrian's walking path spans 7.5 meters, consisting of 4.5 meters from the initial point to the intersection, a street width of 1.5 meters, and an additional 1.5 meters from the crossing to the exit.
On the other hand, the APMV travels a route totaling 8.25 meters, which includes 4.75 meters from the starting position to the intersection, a crossing width of 2 meters, and 1.5 meters from the crossing to the end position.

A baffle plate is positioned near the pedestrians' starting location. 
This plate is structured to block the pedestrians' view, ensuring that they encounter the APMV from a consistent position only after they begin walking.

\subsection{Driving Behaviors of the APMV}

Considering the size of the experimental site, as well as the APMV's maximum speed and acceleration capabilities, the velocity profile of the APMV is depicted by the red line in Fig.~\ref{fig:scenarios}.
The APMV departs from the start position 4.75 meters from the intersection stop line, reaching its maximum speed of 1.67 m/s (comparable to brisk walking) after traveling 1.5 meters. 
The APMV maintains its maximum speed for the next 1.75 meters, then begins to decelerate, and comes to a full stop after covering an additional 1.5 meters.
After the pedestrian has completely crossed the road, the APMV starts moving again and drives to the goal position.
In all trials, the APMV's velocity profiles are the same, regardless of variations in eHMI conditions.

\subsection{eHMI Conditions}
\label{sec:conditions}

This experiment designed four different eHMI conditions (see the right part of Fig.~\ref{fig:scenarios}) to measure a wide range of paired psychological and behavioral data from pedestrians during their interactions with the APMV.
It is assumed that the eHMI cue under each condition, in conjunction with the APMV's driving behaviors, can help pedestrians to vary degrees in understanding the APMV's driving intentions and predicting its driving behavior.
Consequently, this may lead to differences in pedestrians' subjective evaluations and walking behaviors.
Moreover, the designed eHMIs incorporate both visual and voice cues to prevent differences in pedestrians' perceptions of the eHMI information and the timing due to inconsistent attention.
These conditions are detailed in Fig.~\ref{fig:scenarios}.

\subsubsection{Non eHMI Condition}
Participants are required to depart at the same time as APMV.
In encounters between pedestrians and APMVs, the eHMI neither shows information on the screen nor produces sound. 
Consequently, the eHMI does not assist pedestrians in comprehending the APMV's driving intentions and predicting its behavior.
This aligns with existing traffic scenarios where pedestrians only rely solely on vehicle kinematics (\eg distance, speed and acceleration) to infer its driving intentions.

\subsubsection{Early eHMI Condition}
Participants are required to depart at the same time as APMV.
After the APMV departs and accelerates for 1.5 meters, reaching its maximum speed of 1.67 m/s, it displays ``I will stop'' on the eHMI screen and provides a voice cue
(see Fig.~\ref{fig:scenarios}). 
After that, the eHMI will continuously display "I will stop" until the APMV stops in front of the stop line.
After stopping, the eHMI will be turned off.
According to the situation awareness model~\citep{endsley2000} (see Fig.~\ref{fig:model}), we assumed that this early cue may support pedestrians to understand the APMV's driving intentions, but may remain difficult in the real-time prediction of driving behaviors, potentially affecting pedestrians' trust in the APMV and their decision-making.

\subsubsection{Sync eHMI Condition}
Participants are required to depart at the same time as APMV.
The APMV begins to decelerate at a distance of 1.5 meters before reaching the stop line.
At this moment, the eHMI issues visual and voice cues, stating ``I will stop'' (see Fig.~\ref{fig:scenarios}).
At the stopping line, the eHMI will be turned off.
Since Sync eHMI provides information cues that are synchronously aligned with APMV driving behavior (\ie achieving verbal-action synchrony), it may support both the comprehension of the APMV's driving intentions and projection of its future behavior within pedestrians' situation awareness~\citep{endsley2000} (see Fig.~\ref{fig:model}). 
This verbal-action synchrony may further facilitate more accurate risk evaluation and more adaptive decision-making by pedestrians.

\subsubsection{Late eHMI Condition}
Participants are required to depart at the same time as APMV.
During the process from the APMV's departure to its stop, the eHMI does not provide any information cues to pedestrians.
When the APMV stopped, the eHMI will provide ``I stopped'' using both visual and voice cues (see Fig.~\ref{fig:scenarios}).
The visual cue will continue until the APMV departs after the pedestrian completes crossing the road.
Since the Late eHMI does not provide any cues prior to the APMV's stop, pedestrians can infer its driving intentions based only on the vehicle's kinematics.
Once the APMV stops and provides ``I stopped'' via the eHMI, this serves as a post hoc confirmation of driving intention. 
This confirmation information may help pedestrians understand that the APMV has already stopped, reducing their perceived risk and enhancing their trust in the APMV, after it stopped.

\subsection{Experimental Procedure}

At first, the participants were briefed on the experiment, covering details about the APMV's hardware and its self-driving capabilities. 
Then, examples were used to familiarize them with the eHMI, demonstrating its visual and voice cues.
However, details on the eHMI conditions and their activation timing were not provided.
The participants provided their informed consent after all their questions were answered and then began the experiment.

Since all participants reported having no prior encounter experience with the APMV in their lives, we placed the \textit{Non eHMI} condition at the beginning of the experience order to familiarize them with the interaction and establish a baseline for comparison.
The remaining three eHMI conditions provide six potential sequence combinations using a Latin square design.
As detailed in Table~\ref{tab:order}, each experience order is randomly assigned to seven participants to minimize the potential impact of the order effects on experimental results.
Each eHMI condition will be carried out continuously for three trials to obtain stable subjective evaluations and behavioral data.

\begin{figure*}[ht]
\centering
\captionof{table}{eHMI experience orders for the 42 participants. }
\label{tab:order}
\renewcommand{\arraystretch}{1}
\setlength\tabcolsep{1.5pt}
\footnotesize
\begin{tabular}{@{}crcrcrcrcrcrcrcr@{}}
\toprule
Participants & \multicolumn{15}{c}{Order of experience with eHMI conditions} \\ \midrule
(N=42) & \multicolumn{2}{l}{Dummy trial}   & \multicolumn{1}{c}{1st cond.}&  & \multicolumn{2}{l}{Dummy trial}   & \multicolumn{1}{c}{2nd cond.}&  & \multicolumn{2}{l}{Dummy trial}   & \multicolumn{1}{c}{3rd cond.}&  & \multicolumn{2}{l}{Dummy trial}   & \multicolumn{1}{c}{4th cond.} \\ \midrule
N=7 & Non-yielding & $\rightarrow$ & Non eHMI$\times$3 & $\rightarrow$ & Non-yielding & $\rightarrow$ & Early eHMI$\times$3 & $\rightarrow$ & Non-yielding & $\rightarrow$ & Sync eHMI$\times$3 & $\rightarrow$ & Non-yielding & $\rightarrow$ & Late eHMI$\times$3 \\
N=7 & Non-yielding & $\rightarrow$ & Non eHMI$\times$3 & $\rightarrow$ & Non-yielding & $\rightarrow$ & Early eHMI$\times$3 & $\rightarrow$ & Non-yielding & $\rightarrow$ & Late eHMI$\times$3 & $\rightarrow$ & Non-yielding & $\rightarrow$ & Sync eHMI$\times$3 \\
N=7 & Non-yielding & $\rightarrow$ & Non eHMI$\times$3 & $\rightarrow$ & Non-yielding & $\rightarrow$ & Sync eHMI$\times$3 & $\rightarrow$ & Non-yielding & $\rightarrow$ & Early eHMI$\times$3 & $\rightarrow$ & Non-yielding & $\rightarrow$ & Late eHMI$\times$3 \\
N=7 & Non-yielding & $\rightarrow$ & Non eHMI$\times$3 & $\rightarrow$ & Non-yielding & $\rightarrow$ & Sync eHMI$\times$3 & $\rightarrow$ & Non-yielding & $\rightarrow$ & Late eHMI$\times$3 & $\rightarrow$ & Non-yielding & $\rightarrow$ & Early eHMI$\times$3 \\
N=7 & Non-yielding & $\rightarrow$ & Non eHMI$\times$3 & $\rightarrow$ & Non-yielding & $\rightarrow$ & Late eHMI$\times$3 & $\rightarrow$ & Non-yielding & $\rightarrow$ & Early eHMI$\times$3 & $\rightarrow$ & Non-yielding & $\rightarrow$ & Sync eHMI$\times$3 \\
N=7 & Non-yielding & $\rightarrow$ & Non eHMI$\times$3 & $\rightarrow$ & Non-yielding & $\rightarrow$ & Late eHMI$\times$3 & $\rightarrow$ & Non-yielding & $\rightarrow$ & Sync eHMI$\times$3 & $\rightarrow$ & Non-yielding & $\rightarrow$ & Early eHMI$\times$3 \\ \bottomrule
\end{tabular}
\end{figure*}

Before each eHMI condition, a non-yielding dummy trial is performed to remind participants that the APMV might not always stop.
The APMV departs one second before the participants.
It crosses the intersection without decelerating or stopping because pedestrians are about 1 to 1.5 meters from the intersection's edge when it passes the stop line.
The eHMI does not show any information during this process. 

As prior instructions, participants were informed that during the experiment, you would encounter an APMV equipped with various types of eHMIs, and these eHMIs are randomly used in each trial.
Participants were also informed that the eHMI could simultaneously convey both visual and voice cues but the specific configurations of each eHMI condition were not instruct to them.
Finally, participants were given false information that the APMV would automatically decide whether to yield to them based on their distance and movement speed.

In each trial, participants were instructed to walk naturally from the initial point, cross the road, and exit through a door (refer to the blue line in Fig.~\ref{fig:scenarios}).
During this walking process, participants needed to decide whether to yield to the APMV, just as they would in real traffic.
After completing each trial, participants were required to complete a post-trial questionnaire at the desk outside the door.
Then, participants returned to the starting point to continue the next trial.

\subsection{Measurements}
\label{sec:Measurements}

\subsubsection{Post-trial Questioners}
\label{sec:Q}

Based on the hypothesized model shown in Fig.~\ref{fig:model}, \citet{liu2021importance} designed the following six questions to evaluate each step of this model.
This study uses these six questions to evaluate pedestrians' subjective experiences during their interaction with the APMV, both when approaching and crossing the intersection.
They~\footnote{Since the questionnaire was completed after each trial, Q1–Q6 were phrased in the past tense to reflect participants' experiences during the interaction with the APMV, not their post-crossing feelings. This was clearly explained in the pre-experiment instructions.} are:
\begin{itemize}
\small{
\itemsep 0.3em
    \item[Q1:] It was easy to understand the driving intentions of the APMV.
    \item[Q2:] It was easy to predict the driving behaviors of the APMV.
    \item[Q3:] I felt it was dangerous to cross the road when I encountered the APMV.
    \item[Q4:] I trusted the APMV to interact with me safely when I crossed the road.
    \item[Q5:] I felt a sense of relief when I crossed the road.
    \item[Q6:] I felt hesitant to make the decision of crossing the road or not when I encountered the APMV.
    }
\end{itemize}
These six questions were required to be answered using a 5-point Likert scale, \ie 1=``strongly disagree'', 2=``disagree'', 3=``neutral'', 4=``agree'', and 5=``strongly agree''.

Q1 and Q2 assess comprehension and projection in pedestrian situation awareness during the interaction. 
Q3, Q4, and Q5 evaluate the risk homeostasis process.
Q6 measures hesitation in decision-making about crossing the road.
Q1, Q2, Q3, and Q6 are designed to measure pedestrians' subjective experiences throughout the entire interaction process, from initially perceiving the APMV to exiting the room.
Moreover, although Q4 and Q5 are phrased to assess participants' subjective experiences during the crossing, these questions are designed to capture the trust in APMVs and the sense of relief that pedestrians had accumulated before the crossing and during the crossing.
This is because trust is a cumulative process that develops over time~\citep{hu2021trust, walker2023trust}, and the sense of relief depends on the level of trust~\citep{Matsubayashi_2023}.
Therefore, the responses of Q4 and Q5 not only represent the states experienced during the crossing, but also capture whether sufficient trust and relief had already been built beforehand.

Note that, based on the four dimensions of trust in human-machine systems proposed by~\citet{lee1992trust}, \ie foundation, purpose, process, and performance, we considered that performance-based trust is most important for pedestrians interacting with an APMV. 
This refers to pedestrians' trust in the APMV's performance to avoid collisions and ensure safety for pedestrians.
Thus, Q4 was designed to capture this trust.

\subsubsection{Walking Durations in Three Crossing Phases}
\label{sec:Walking times}

Walking durations of the participants are calculated in the three crossing phases.
Crossing Initiation Time (CIT) refers to the time it takes for pedestrians to decide to cross the road.
This metric was used to indicate the interval between the moment pedestrians initially perceive the APMV and the moment they step into the intersection.
Crossing Time~(CT) refers to the duration between the moment pedestrians enter the intersection and the moment they completely cross the intersection.
After Crossing Time (ACT) refers to the time it takes pedestrians to walk from the end of the intersection to the designated end point.

\begin{figure}[!t]
  \centering
  \includegraphics[width=1\linewidth,trim=5 10 5 23,clip]{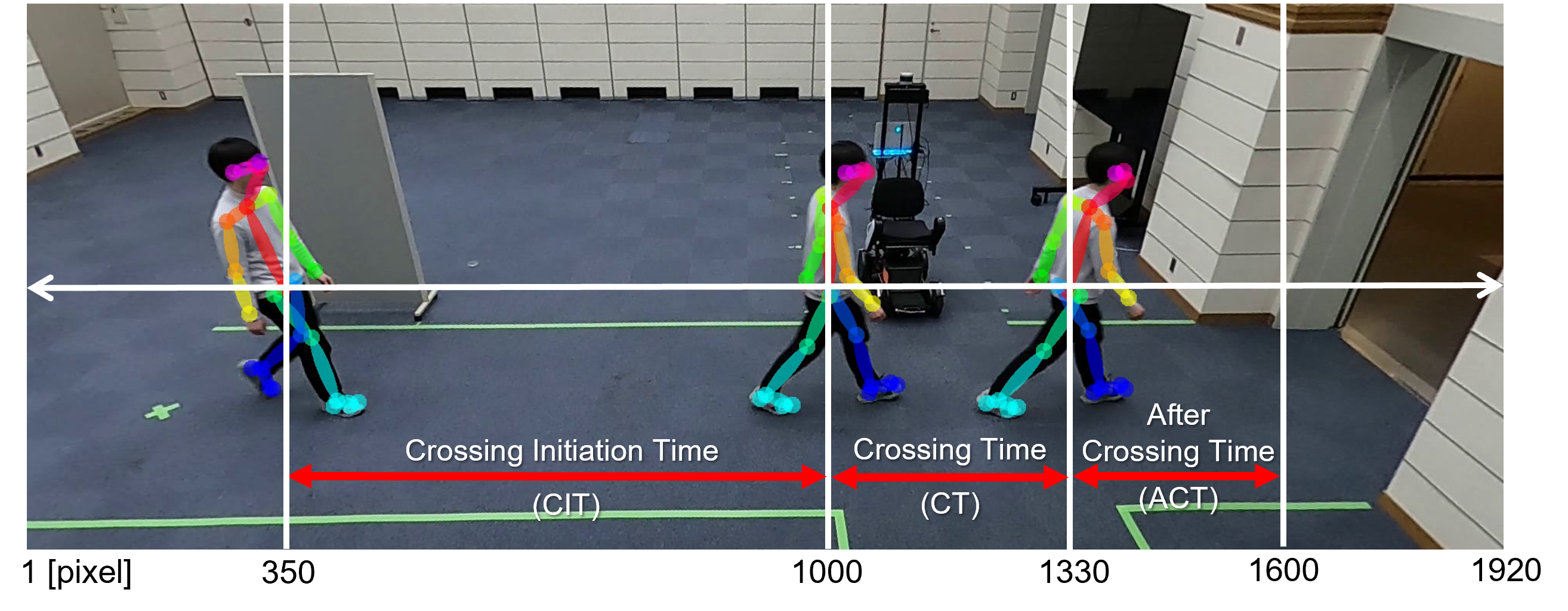}
  \caption{Participants' walking durations are measured in three phases: crossing initiation time (CIT), crossing time (CT), and after crossing time (ACT), using the MidHip keypoint estimated by OpenPose with the BODY~25 joint set.}
   \label{fig:OpenPose} 
\end{figure}

To measure the walking durations mentioned above, the walking processes of participants were recorded by a camera (1920$\times$1080 pixels with 30 FPS) shown in Fig.~\ref{fig:scenarios}.
Then, OpenPose~\citep{openPose} based on the BODY~25 joint set is used to extract skeletal features of participants during their walking process (see Fig.~\ref{fig:OpenPose}).
Then, the feature point of the waist called the MidHip point is used to represent the walking position of participants in the image coordinate system.
The existence time of the MidHip point during the three stages is used to represent CIT, CT, and ACT.
Specifically, as shown in Fig.~\ref{fig:OpenPose}, CIT is defined as the time taken for the pedestrian's MidHip keypoint to move from 350 pixels to 1000 pixels;
CT is the time taken for the MidHip keypoint to move from 1000 pixels to 1330 pixels; while ACT is the time taken for it to move from 1330 pixels to 1600 pixels on the image's horizontal axis.

\begin{table*}[h!b]
\centering
\caption{Descriptive statistics of all factors under four eHMI conditions and their Spearman correlation coefficients. }
 \label{tab:correlation}
\footnotesize
\setlength\tabcolsep{1.7pt}
\renewcommand{\arraystretch}{1}
\begin{tabular}{lcccccccrrrrrrrrrr}
\toprule
\multicolumn{1}{c}{\multirow{2}{*}{Factor}}&  \multicolumn{7}{c}{Descriptive Statistics}& \multicolumn{9}{c}{Spearman Correlation Coefficient} & \multicolumn{1}{r}{\multirow{2}{*}{VIF}}\\ \cmidrule(lr){2-8} \cmidrule(l){9-17} 
  & \multicolumn{1}{c}{\textit{N}}& \multicolumn{1}{c}{mean}& \multicolumn{1}{c}{std.}&   median&IQR&\multicolumn{1}{c}{min}& \multicolumn{1}{c}{max}& \multicolumn{1}{c}{Q1}& \multicolumn{1}{c}{Q2}& \multicolumn{1}{c}{Q3}& \multicolumn{1}{c}{Q4}& \multicolumn{1}{c}{Q5}& \multicolumn{1}{c}{Q6}& \multicolumn{1}{c}{CIT}& \multicolumn{1}{c}{CT}& \multicolumn{1}{c}{ACT} & \\  \cmidrule(lr){1-8} \cmidrule(l){9-17}  \cmidrule(l){18-18} 
Q1 Understand  & 504 & 3.97 & 1.20 &   4.00&2.00&1.00 & 5.00  &\gradient{1.00}***&\gradient{0.78}***&\gradient{-0.67}***&\gradient{0.68}***&\gradient{0.42}***&\gradient{-0.70}***&\gradient{-0.32}***&\gradient{-0.11}*~ ~&\gradient{-0.08}~ ~ ~&3.53\\
Q2 Predict & 504 & 3.89 & 1.20 &   4.00&2.00&1.00 & 5.00 &\gradient{0.78}***&\gradient{1.00}***&\gradient{-0.68}***&\gradient{0.66}***&\gradient{0.41}***&\gradient{-0.69}***&\gradient{-0.28}***&\gradient{-0.12}**~ &\gradient{-0.09}*~ ~ &3.68\\
Q3 Dangerous & 504 & 2.17 & 1.20 &   2.00&2.00&1.00 & 5.00  &\gradient{-0.67}***&\gradient{-0.68}***&\gradient{1.00}***&\gradient{-0.75}***&\gradient{-0.46}***&\gradient{0.77}***&\gradient{0.19}***&\gradient{0.03}~ ~ ~&\gradient{0.02}~ ~ ~&3.25\\
Q4 Trust & 504 & 3.81 & 1.16 &   4.00&2.00&1.00 & 5.00 &\gradient{0.68}***&\gradient{0.66}***&\gradient{-0.75}***&\gradient{1.00}***&\gradient{0.59}***&\gradient{-0.72}***&\gradient{-0.25}***&\gradient{-0.05}~ ~ ~&\gradient{0.00}~ ~ ~&2.94\\
Q5 Relief & 504 & 3.81 & 1.17 &   4.00&2.00&1.00 & 5.00 &\gradient{0.42}***&\gradient{0.41}***&\gradient{-0.46}***&\gradient{0.59}***&\gradient{1.00}***&\gradient{-0.43}***&\gradient{-0.25}***&\gradient{-0.07}~ ~ ~&\gradient{-0.02}~ ~ ~&1.37\\
Q6 Hesitant& 504 & 2.51 & 1.35 &   2.00&3.00&1.00 & 5.00 &\gradient{-0.70}***&\gradient{-0.69}***&\gradient{0.77}***&\gradient{-0.72}***&\gradient{-0.43}***&\gradient{1.00}***&\gradient{0.30}***&\gradient{0.08} ~ ~ &\gradient{0.05}~ ~ ~&3.33\\
CIT  & 504 & 3.06 & 0.83 &   2.87&1.00&1.70 & 6.11  &\gradient{-0.32}***&\gradient{-0.28}***&\gradient{0.19}***&\gradient{-0.25}***&\gradient{-0.25}***&\gradient{0.30}***&\gradient{1.00}***&\gradient{0.45}***&\gradient{0.44}***&1.40\\
CT  & 504 & 1.19 & 0.18 &   1.17&0.20&0.73 & 2.37  &\gradient{-0.11}*~ ~ &\gradient{-0.12}**~ &\gradient{0.03}~ ~ ~&\gradient{-0.05}~ ~ ~&\gradient{-0.07}~ ~ ~&\gradient{0.08}~ ~ ~&\gradient{0.45}***&\gradient{1.00}***&\gradient{0.80}***&2.52\\
ACT  & 504 & 0.96 & 0.13 &   0.93&0.17&0.60 & 1.33 &\gradient{-0.08}~ ~ ~&\gradient{-0.09}*~ ~ &\gradient{0.02}~ ~ ~&\gradient{0.00}~ ~ ~&\gradient{-0.02}~ ~ ~&\gradient{0.05}~ ~ ~&\gradient{0.44}***&\gradient{0.80}***&\gradient{1.00}***&2.53\\ \bottomrule
\multicolumn{18}{l}{ \footnotesize \renewcommand{\arraystretch}{1} \begin{tabular}[c]{@{}l@{}}Q1 to Q6: 1=``strongly disagree'', 2=``disagree'', 3=``neutral'', 4=``agree'', 5=``strongly agree''. CIT, CT and ACT: the unit is in second.\\ $*:p<.05, **:p<.01, ***:p<.001$.  VIF: Variance Inflation Factor (commonly acceptable threshold: VIF$<4$ or VIF$<10$~\citep{o2007caution}).\end{tabular}}
\end{tabular}

\end{table*}

\subsection{Prior Knowledge of DirectLiNGAM}

After measuring the data on participants' subjective evaluations and walking behaviors, we applied DirectLiNGAM to discover the causal relations from their subjective evaluations to walking behaviors.
Based on the hypothesized model shown in Fig.~\ref{fig:model}, we have set prior knowledge in DirectLiNGAM to specify the inputs and outputs of this process. 
Specifically, although perception is the first stage of situation awareness, it refers to the behavioral process of detecting or noticing elements in the environment and primarily involves sensory input, such as visual or auditory signals.
Therefore, Q1 is designated as input, \ie a prior exogenous variable, because comprehension could be considered as the initial stage of pedestrians' cognition (rather than awareness).
This means that Q1 is assumed to be an independent factor within the model, unaffected by other variables but capable of influencing them.
Furthermore, as the aim of this study is to discover the causal relations from pedestrians' psychological states to their behavior, CIT, CT, and ACT are designated as the outputs, \ie prior endogenous variables.
This means they are assumed to be dependent variables,  influenced by other variables but do not influence others.
The causal relations among the remaining variables are inferred using DirectLiNGAM.

\section{RESULTS}
\subsection{Data Summary}
Considering the differences in interaction timing and distance between pedestrians and the APMV in the non-yielding dummy condition compared to other eHMI conditions, data from this condition is excluded from the analysis.

Throughout the experiment, data from a total of 504 trials (126 trials per eHMI condition) were collected, each consisting of nine factors: six subjective evaluation factors (Q1 to Q6) and three factors of walking behavior (CIT, CT and ACT).
Table~\ref{tab:correlation} shows the descriptive statistics of each factor under all eHMI conditions.
By assessing the maximum, minimum values, and standard deviation (std) for each factor, the collected data are deemed comprehensive, covering a wide range of psychological states and walking behaviors of pedestrians, which is suitable for causal discovery.
Table~\ref{tab:correlation} also shows Spearman correlation coefficients between each pair of factors, which helps us to preliminarily discover potential relations among them.

To assess multicollinearity among the independent variables, we computed the Variance Inflation Factor (VIF)~\citep{o2007caution} for each factor. 
Table~\ref{tab:correlation} also presents the VIF for each factor.
According to a commonly acceptable threshold of VIF below 4~\citep{o2007caution}, the multicollinearity of each factor was considered acceptable.

\subsection{Subjective Evaluations and Walking Behaviors under each eHMI Condition}

\begin{figure}[b]
   \centering
  \includegraphics[width=1\linewidth]{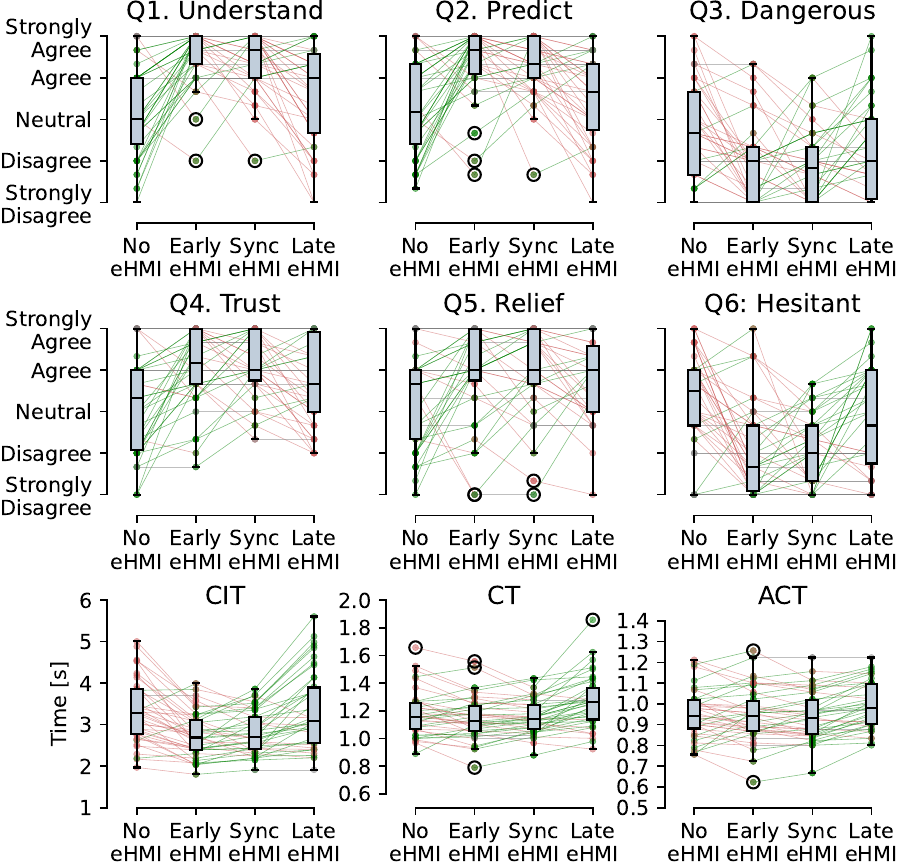}  
  \captionof{figure}{Results of subjective evaluations (Q1 to Q6) and walking behavior factors (CIT, CT and ACT).}
  \label{fig:results}
\end{figure}

Under each eHMI condition, the results of subjective evaluations and walking behavior factors are presented in Fig.~\ref{fig:results} as box plots with paired point plots.
These box plots visualize a comparison of the median and distribution of responses under the different eHMI conditions.
These paired-point plots show average factor values for each participant in three trials per eHMI condition.
Lines between conditions indicate changes in the relevant factors for each participant. 
The green line shows higher average factor values on the right condition, while the red line shows the opposite.

\begin{figure}[!hb]
\captionof{table}{Friedman test results for subjective evaluations and walking behavior factor under four eHMI conditions.}
\label{tab:friedman}
\footnotesize
\setlength\tabcolsep{8pt}.
\renewcommand{\arraystretch}{0.9}
\centering
\begin{tabular}{@{}lcrrrr@{}}
\toprule
Factor & \textit{W}& \multicolumn{1}{c}{ddof1}& \multicolumn{1}{c}{ddof2}& \multicolumn{1}{c}{\textit{F}}& \multicolumn{1}{c}{\textit{p}} \\ \midrule
Q1 Understand  
& 0.407 & 2.952& 121.048& 28.098 & \textless{}.001~***\\
Q2 Predict 
& 0.357 & 2.952& 121.048& 22.778 & \textless{}.001~***\\
Q3 Dangerous 
& 0.195 & 2.952& 121.048& 9.948 & \textless{}.001~***\\
Q4 Trust 
& 0.248 & 2.952& 121.048& 13.542 & \textless{}.001~***\\
Q5 Relief 
& 0.118 & 2.952& 121.048& 5.511 & \textless{}.001~***\\
Q6 Hesitant& 0.232 & 2.952& 121.048& 12.359 &  \textless{}.001~***\\ 
CIT& 0.423 & 2.952& 121.048& 30.075 & \textless{}.001~***\\
CT & 0.257 & 2.952& 121.048& 14.212 &  \textless{}.001~***\\
ACT& 0.188 & 2.952&  121.048& 9.468 &  \textless{}.001~***\\ \bottomrule
\multicolumn{6}{l}{ \footnotesize \renewcommand{\arraystretch}{1} \begin{tabular}[c]{@{}l@{}}***:$p<.001$.\end{tabular}}
\end{tabular}
\end{figure}

\begin{figure}[!hb]
\renewcommand{\arraystretch}{0.8}
\setlength\tabcolsep{3pt}
\footnotesize
\centering
\captionof{table}{Post-hoc two-sided pairwise comparisons using the Wilcoxon signed-rank test with Benjamini/Hochberg FDR correction for the main effect of eHMI conditions on the factors of subjective evaluations and walking behavior.}
\label{tab:post-hoc}
\begin{tabular}{@{}llcccccrr@{\hspace{1pt}}lc@{}}
\toprule
 &
  \begin{tabular}[c]{@{}c@{}}eHMI\\ A\end{tabular} &
  \begin{tabular}[c]{@{}c@{}}eHMI\\ B\end{tabular} &
  \begin{tabular}[c]{@{}c@{}}Mean\\ (A)\end{tabular} &
  \begin{tabular}[c]{@{}c@{}}Std.\\ (A)\end{tabular} &
  \begin{tabular}[c]{@{}c@{}}Mean\\ (B)\end{tabular} &
  \begin{tabular}[c]{@{}c@{}}Std.\\ (B)\end{tabular} &
  \multicolumn{1}{c}{\textit{W}} &
  \multicolumn{2}{c}{\textit{p}-adj} &
  CLES \\ \midrule
\multirow{6}{*}{Q1}  & Early & Late & 4.57 & 0.64 & 3.59 & 1.21 & 42.0  & \textless{}.001 & *** & 0.757 \\
                     & Early & Non  & 4.57 & 0.64 & 3.28 & 1.20 & 46.0  & \textless{}.001 & *** & 0.816 \\
                     & Early & Sync & 4.57 & 0.64 & 4.43 & 0.69 & 76.0  & .122            &     & 0.560 \\
                     & Late  & Non  & 3.59 & 1.21 & 3.28 & 1.20 & 188.0 & .157            &     & 0.579 \\
                     & Late  & Sync & 3.59 & 1.21 & 4.43 & 0.69 & 58.5  & \textless{}.001 & *** & 0.291 \\
                     & Non   & Sync & 3.28 & 1.20 & 4.43 & 0.69 & 15.5  & \textless{}.001 & *** & 0.216 \\ \midrule
\multirow{6}{*}{Q2}  & Early & Late & 4.44 & 0.81 & 3.54 & 1.10 & 74.0  & \textless{}.001 & *** & 0.762 \\
                     & Early & Non  & 4.44 & 0.81 & 3.30 & 1.16 & 81.0  & \textless{}.001 & *** & 0.788 \\
                     & Early & Sync & 4.44 & 0.81 & 4.29 & 0.71 & 128.0 & .145            &     & 0.592 \\
                     & Late  & Non  & 3.54 & 1.10 & 3.30 & 1.16 & 195.5 & .099            &     & 0.562 \\
                     & Late  & Sync & 3.54 & 1.10 & 4.29 & 0.71 & 61.5  & \textless{}.001 & *** & 0.297 \\
                     & Non   & Sync & 3.30 & 1.16 & 4.29 & 0.71 & 55.0  & \textless{}.001 & *** & 0.253 \\ \midrule
\multirow{6}{*}{Q3}  & Early & Late & 1.91 & 0.97 & 2.25 & 1.11 & 121.0 & .076            &     & 0.411 \\
                     & Early & Non  & 1.91 & 0.97 & 2.68 & 1.23 & 84.5  & .002            & **  & 0.312 \\
                     & Early & Sync & 1.91 & 0.97 & 1.85 & 0.86 & 156.0 & .434            &     & 0.507 \\
                     & Late  & Non  & 2.25 & 1.11 & 2.68 & 1.23 & 156.0 & .066            &     & 0.400 \\
                     & Late  & Sync & 2.25 & 1.11 & 1.85 & 0.86 & 119.0 & .040            & *   & 0.600 \\
                     & Non   & Sync & 2.68 & 1.23 & 1.85 & 0.86 & 55.0  & \textless{}.001 & *** & 0.700 \\ \midrule
\multirow{6}{*}{Q4}  & Early & Late & 4.09 & 0.90 & 3.73 & 1.02 & 139.0 & .040            & *   & 0.598 \\
                     & Early & Non  & 4.09 & 0.90 & 3.25 & 1.23 & 51.5  & \textless{}.001 & *** & 0.689 \\
                     & Early & Sync & 4.09 & 0.90 & 4.17 & 0.77 & 156.0 & .629            &     & 0.480 \\
                     & Late  & Non  & 3.73 & 1.02 & 3.25 & 1.23 & 75.5  & .010            & *   & 0.613 \\
                     & Late  & Sync & 3.73 & 1.02 & 4.17 & 0.77 & 71.0  & .005            & **  & 0.381 \\
                     & Non   & Sync & 3.25 & 1.23 & 4.17 & 0.77 & 25.5  & \textless{}.001 & *** & 0.285 \\ \midrule
\multirow{6}{*}{Q5}  & Early & Late & 3.98 & 1.08 & 3.82 & 0.96 & 105.0 & .149            &     & 0.562 \\
                     & Early & Non  & 3.98 & 1.08 & 3.41 & 1.21 & 86.0  & .022            & *   & 0.642 \\
                     & Early & Sync & 3.98 & 1.08 & 4.05 & 1.04 & 156.5 & .637            &     & 0.489 \\
                     & Late  & Non  & 3.82 & 0.96 & 3.41 & 1.21 & 133.0 & .049            & *   & 0.598 \\
                     & Late  & Sync & 3.82 & 0.96 & 4.05 & 1.04 & 99.5  & .082            &     & 0.422 \\
                     & Non   & Sync & 3.41 & 1.21 & 4.05 & 1.04 & 120.0 & .022            & *   & 0.351 \\ \midrule
\multirow{6}{*}{Q6}  & Early & Late & 2.07 & 1.08 & 2.72 & 1.26 & 96.5  & .005            & **  & 0.348 \\
                     & Early & Non  & 2.07 & 1.08 & 3.23 & 1.24 & 57.5  & \textless{}.001 & *** & 0.249 \\
                     & Early & Sync & 2.07 & 1.08 & 2.02 & 0.83 & 195.0 & .864            &     & 0.483 \\
                     & Late  & Non  & 2.72 & 1.26 & 3.23 & 1.24 & 143.5 & .017            & *   & 0.389 \\
                     & Late  & Sync & 2.72 & 1.26 & 2.02 & 0.83 & 52.0  & \textless{}.001 & *** & 0.654 \\
                     & Non   & Sync & 3.23 & 1.24 & 2.02 & 0.83 & 17.5  & \textless{}.001 & *** & 0.780 \\ \midrule
\multirow{6}{*}{CIT} & Early & Late & 2.76 & 0.53 & 3.34 & 0.94 & 56.0  & \textless{}.001 & *** & 0.322 \\
                     & Early & Non  & 2.76 & 0.53 & 3.32 & 0.76 & 54.0  & \textless{}.001 & *** & 0.277 \\
                     & Early & Sync & 2.76 & 0.53 & 2.82 & 0.51 & 339.0 & .196            &     & 0.470 \\
                     & Late  & Non  & 3.34 & 0.94 & 3.32 & 0.76 & 413.5 & .831            &     & 0.484 \\
                     & Late  & Sync & 3.34 & 0.94 & 2.82 & 0.51 & 89.5  & \textless{}.001 & *** & 0.656 \\
                     & Non   & Sync & 3.32 & 0.76 & 2.82 & 0.51 & 55.0  & \textless{}.001 & *** & 0.694 \\ \midrule
\multirow{6}{*}{CT}  & Early & Late & 1.15 & 0.15 & 1.26 & 0.19 & 60.5  & \textless{}.001 & *** & 0.312 \\
                     & Early & Non  & 1.15 & 0.15 & 1.18 & 0.16 & 335.5 & .228            &     & 0.471 \\
                     & Early & Sync & 1.15 & 0.15 & 1.16 & 0.12 & 323.5 & .247            &     & 0.481 \\
                     & Late  & Non  & 1.26 & 0.19 & 1.18 & 0.16 & 133.5 & \textless{}.001 & *** & 0.657 \\
                     & Late  & Sync & 1.26 & 0.19 & 1.16 & 0.12 & 59.5  & \textless{}.001 & *** & 0.677 \\
                     & Non   & Sync & 1.18 & 0.16 & 1.16 & 0.12 & 330.5 & .237            &     & 0.516 \\ \midrule
\multirow{6}{*}{ACT} & Early & Late & 0.95 & 0.13 & 0.99 & 0.11 & 83.0  & \textless{}.001 & *** & 0.390 \\
                     & Early & Non  & 0.95 & 0.13 & 0.95 & 0.12 & 389.5 & .539            &     & 0.478 \\
                     & Early & Sync & 0.95 & 0.13 & 0.94 & 0.12 & 348.0 & .562            &     & 0.523 \\
                     & Late  & Non  & 0.99 & 0.11 & 0.95 & 0.12 & 204.5 & .007            & **  & 0.585 \\
                     & Late  & Sync & 0.99 & 0.11 & 0.94 & 0.12 & 71.5  & \textless{}.001 & *** & 0.623 \\
                     & Non   & Sync & 0.95 & 0.12 & 0.94 & 0.12 & 358.0 & .526            &     & 0.527 \\ \bottomrule
 \multicolumn{11}{l}{ \footnotesize \renewcommand{\arraystretch}{1} \begin{tabular}[c]{@{}l@{}}*:$p<.05$, **:$p<.01$, ***:$p<.001$. CLES: common language effect size.\end{tabular}}
\end{tabular}
\end{figure}

This experiment aims to use various eHMI conditions to influence pedestrians' subjective evaluations and walking behaviors when interacting with APMVs, facilitating causal analysis among these variables.
To validate this motivation, the Friedman test was performed to determine whether there were differences in each factor among four eHMI conditions, as the Shapiro-Wilk test indicated that the results for each factor did not conform to a normal distribution.
The results of the Friedman test are shown in Table~\ref{tab:friedman} that all factors had significant differences among the four eHMI conditions, respectively.
Then, Wilcoxon signed-rank tests with Benjamini/Hochberg FDR corrections were used as post-hoc pairwise comparisons of the main effect of eHMI conditions on subjective evaluations and walking durations, as shown in Table~\ref{tab:post-hoc}.

\subsection{Direct Causal relations and Their Statistical Reliability}

\begin{figure*}[h!b]
  \centering
  \begin{minipage}[t]{0.48\linewidth} 
  \centering
  \includegraphics[width=1\linewidth]{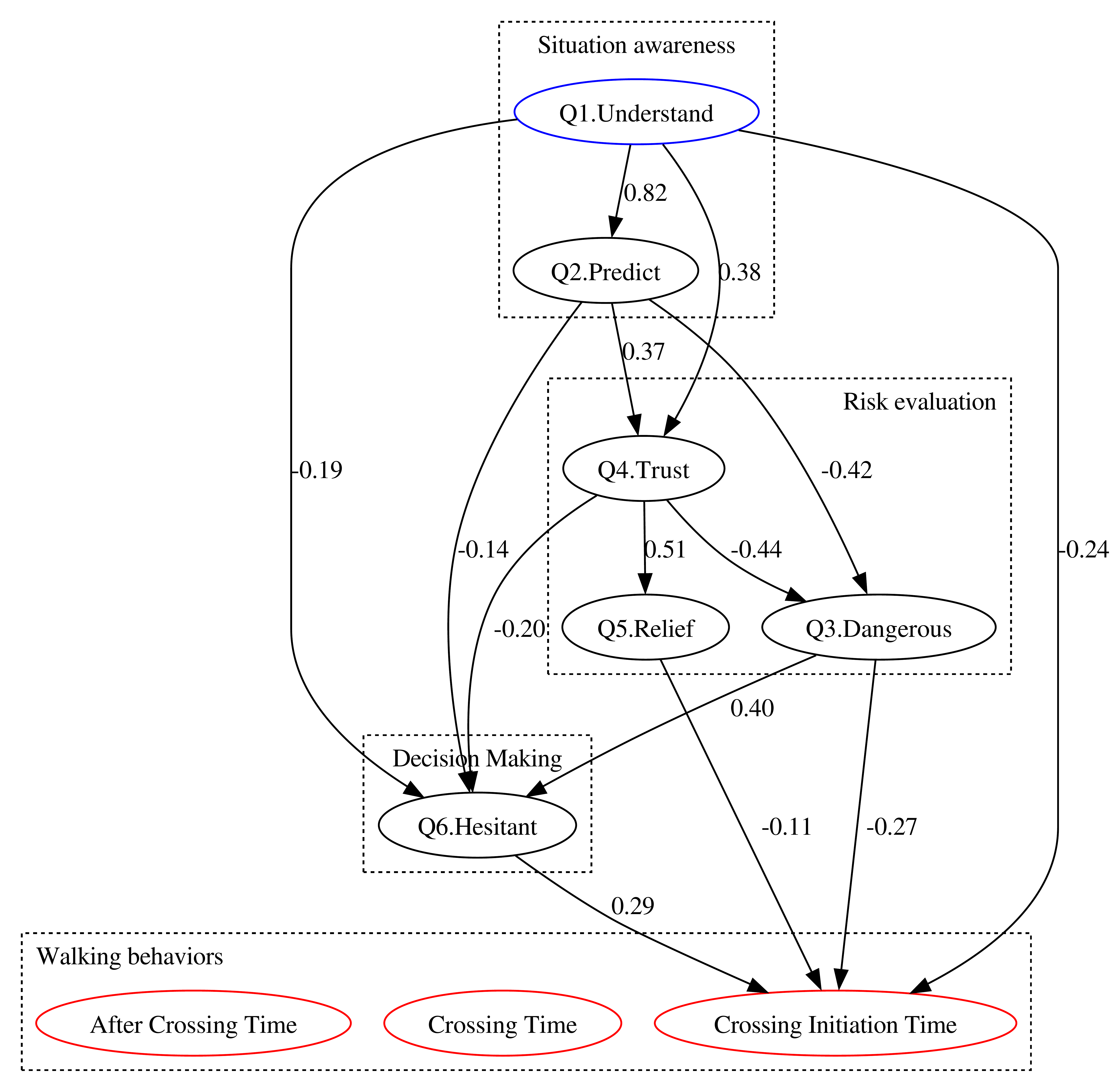}
  \caption{A causal model showing direct causal paths and their corresponding direct causal effects, estimated by DirectLiNGAM from data on nine factors measured across 504 trials. (blue node: a prior exogenous variable, red nodes: prior endogenous variables).}
  \label{fig:LiNGAM}
\end{minipage}
\hspace{2mm}
\centering
 \begin{minipage}[t]{0.48\linewidth} 
\centering
  \includegraphics[width=1\linewidth]{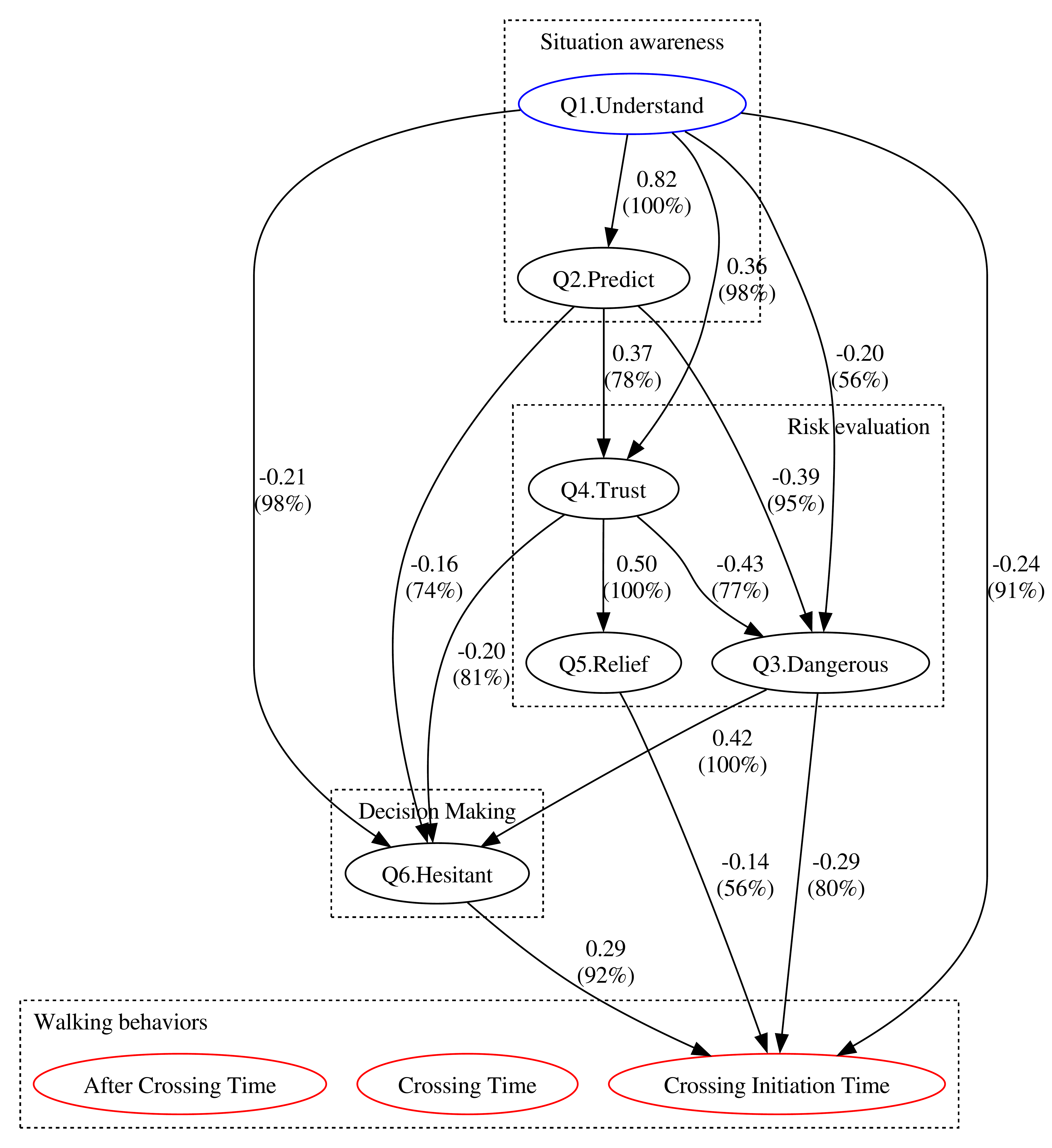}
   \caption{Bootstrapped causal model estimated via DirectLiNGAM 5000 times, showing direct paths with the median direct causal effects and the replication probabilities exceeding 30\%. (blue node: exogenous variable; red node: endogenous variables)}
  \label{fig:Bootstrap}
 \end{minipage}
\end{figure*}

\begin{figure*}[bh]
\centering
\footnotesize
\captionof{table}{Model fit summary of the causal model and the bootstrapped causal model estimated via DirectLiNGAM.}
\label{tab:model_fit}
\setlength\tabcolsep{6pt}
\renewcommand{\arraystretch}{1}
\begin{tabular}{@{}rrrcrrrrrr}
\toprule
& \multicolumn{1}{c}{$\chi^2$ (dof)} & $\textit{p}_{\chi^2}$  & $\chi^2_{baseline}$ (dof) & \multicolumn{1}{c}{CFI} & \multicolumn{1}{c}{GFI} & \multicolumn{1}{c}{AGFI} & \multicolumn{1}{c}{NFI} & \multicolumn{1}{c}{TLI} & \multicolumn{1}{c}{RMSEA}    \\ \midrule
\footnotesize
Causal model (Fig.~\ref{fig:LiNGAM}) & 6.362 (22) & .999 & 2214.372 (22) & 1.000 & 0.997 & 0.997 & 0.997 & 1.007 & 0.000 \\
Bootstrapped causal model (Fig.~\ref{fig:Bootstrap}) & 30.688 (22) & .103 & 2214.372 (22) & 0.996 & 0.986 & 0.986 & 0.986 & 0.996 & 0.028 \\ \midrule
Acceptable thresholds \citep{HOOPER2008} & \multicolumn{1}{c}{-}  & $>.050$ & - & $>0.950$ & $>0.950$ & $>0.950$  & $>0.950$ & $>0.950$ & $<0.070$ \\
\bottomrule
\multicolumn{10}{l}{ \footnotesize \renewcommand{\arraystretch}{1} \begin{tabular}[c]{@{}l@{}}CFI: comparative fit index. GFI: goodness-of-fit index; AGFI: adjusted goodness-of-fit index; NFI: normed fit index. \\ TLI: Tucker-Lewis index; RMSEA: root mean square error of approximation.\end{tabular}}
\end{tabular}
\end{figure*}

\begin{figure*}[t!h]
   \centering
   \includegraphics[width=1\linewidth]{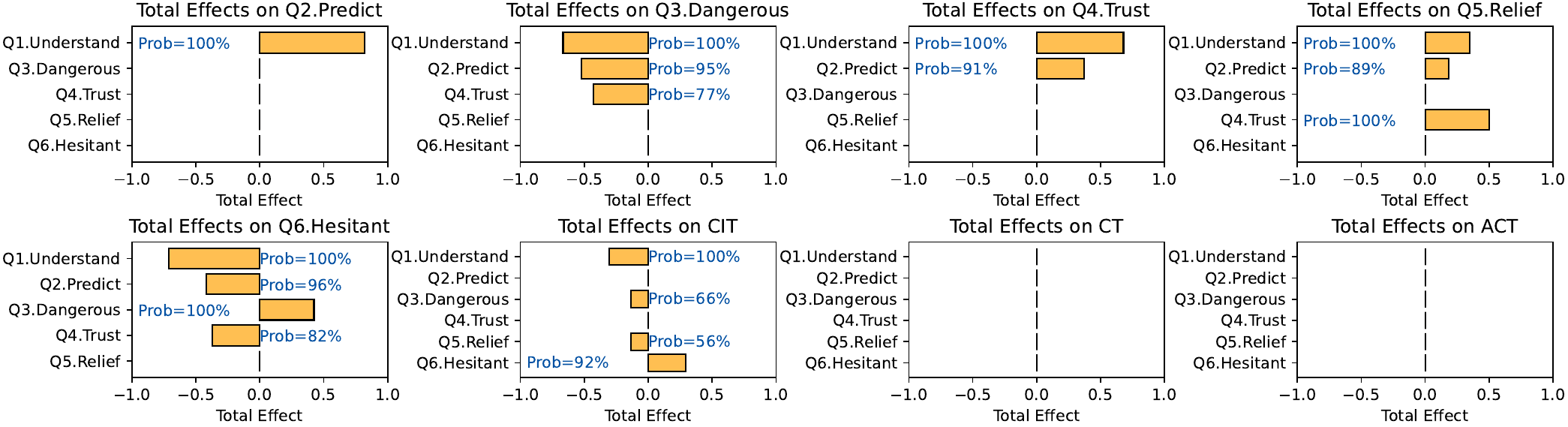}
   \caption{The median total causal effects of 5000 times DirectLiNGAM via bootstrap. The total causal effects with reproducibility probabilities exceeding 30\% are presented.}
   \label{fig:Total_causal_effects}
   
 \end{figure*}

Due to the differences in scale among Q1–Q6, CIT, CT, and ACT, all variables were standardized using z-scores prior to applying DirectLiNGAM.
The estimated causal model using DirectLiNGAM is shown in Fig.~\ref{fig:LiNGAM}.
This causal model shows the direct relations (\ie the direct causal paths and their direct causal effects) among the nine factors.
Arrows between nodes indicate the direct causal path. 
The value next to each arrow represents the direct causal effect corresponding to the elements of the adjacency matrix.

To validate the non-Gaussian error assumption of DirectLiNGAM, Shapiro–Wilk test was separately conducted on the residuals of each variable in the inferred causal model.
The results confirmed that all residuals were significantly non-Gaussian ($p < .001$), which supported the assumption of DirectLiNGAM.

Further, a summary of model fit indices for the causal model estimated using DirectLiNGAM is presented in the first row of Table~\ref{tab:model_fit}. 
The computation formulas and interpretation guidelines for these model fit indices can be found in~\citep{HOOPER2008,schermelleh2003evaluating}.
The result of the chi-square test ($\chi^2(22) = 6.362$, $p_{\chi^2} = .999$) indicated that the causal model provided an acceptable fit. 
In line with the acceptable thresholds of model fit suggested by~\citep{HOOPER2008},
the comparative fit index (CFI), 
goodness-of-fit index (GFI),
adjusted goodness-of-fit index (AGFI), 
normed fit index (NFI), 
Tucker-Lewis index (TLI), 
and root mean square error of approximation (RMSEA)
also indicated that the causal model estimated by DirectLiNGAM fit the data well.
Nonetheless, $p_{\chi^2} = .999, \text{CFI} = 1.000, \text{TLI} = 1.007, \text{RMSEA} = 0.000$ may also imply potential overfitting.

To verify the statistical reliability of the estimated direct causal relations, the bootstrap method was applied~\citep{thamvitayakul2012bootstrap}.
Specifically, bootstrap resampling was conducted to generate new datasets by repeated random sampling from the measured dataset. 
Each resampled dataset retains the same size as the measured dataset.
A total of 5000 times of bootstrap resampling were performed, generating 5000 new datasets.
For each dataset, DirectLiNGAM was used to conduct causal discovery and independently estimate the adjacency matrix $\bm{A}$. 
Thereafter, the 5000 adjacency matrices of $\bm{A}$ were aggregated to calculate the median direct causal effects of non-zero elements and their probability of occurrence, \ie the reproducibility probability of the direct causal relations.
Causal paths with replication probabilities below 30\% were pruned, \ie their corresponding median direct causal effects were set to zero.
Figure~\ref{fig:Bootstrap} shows the estimated median direct causal effects with reproducibility probabilities greater than 30\%, as determined from the 5000 bootstrap with DirectLiNGAM.
The reproducibility probability for each direct causal relation is labeled next to each arrow in Fig.~\ref{fig:Bootstrap}.
The second row of Table~\ref{tab:model_fit} shows a summary of the fit indices for this bootstrap causal model.
The chi-square test result ($\chi^2(22) = 30.688$, $p_{\chi^2} = .103$) indicated that the causal model provided an acceptable fit. 
Additionally, the CFI, GFI, NFI, TLI, and RMSEA values also suggested that the bootstrapped causal model fitted the data well, according to the acceptable thresholds suggested by~\citep{HOOPER2008}.

\subsection{Total Causal Effects and Their Statistical Reliability}

Total causal effect quantifies the change in the effect variable when the cause variable is altered (\ie intervention)~\citep{Bollen_1989}.
It includes both direct causal effects and indirect causal effects from the cause variable to the effect variable.

To calculate a more reliable total effect, the bootstrap method was also used 5000 times.
The bootstrap outputs the medians of total causal effects and their reproducibility probabilities (\ie the probabilities of whether total effects are non-zero) over the DirectLiNGAM results via bootstrap samplings.

Since Q1 is assumed to be an exogenous variable, unaffected by other variables but capable of influencing them, the total causal effects of the other variables on Q2-Q6, as well as on CIT, CT, and ACT, were calculated separately.
These total causal effects with reproducibility probabilities exceeding 30\% are presented in Fig.~\ref{fig:Total_causal_effects}.

\section{DISCUSSION}

\subsection{Causal Discovery from Subjective Evaluations to Walking Behaviors}

The direct causal relations between the factors and the results of their bootstrap-based statistical reliability are shown in Figs.~\ref{fig:LiNGAM} and~\ref{fig:Bootstrap}, respectively.
All the direct causal relations between factors shown in Fig.~\ref{fig:LiNGAM} are confirmed to have at least 56\% reliability as shown in Fig.~\ref{fig:Bootstrap}.
The only difference is that, compared to the causal relation discovered directly from the collected data, a new direct causal relation from Q1 to Q3 is discovered after using 5000 times of bootstrap, with a reliability of 56\%.
This suggests that as the number of participants increases further, understanding the driving intention may directly affect their sense of danger during encounters with the APMV.

Overall, the general flow of the results from causal discovery aligns with the hypothesized model (see Fig. \ref{fig:model}), where the outcomes of situation awareness lead to risk evaluation, the outcomes of risk evaluation lead to hesitation in decision-making, and the results of hesitation lead to walking behaviors.

Next, we will discuss in detail the findings of causal effects on each factor by combining their reliable direct causal effects (see Fig.~\ref{fig:Bootstrap}) and total causal effects (see Fig.~\ref{fig:Total_causal_effects}).

\subsubsection{Causal Effects on the Prediction of APMV's behaviors}

Fig.~\ref{fig:Bootstrap} shows a high reproducibility probability (100\%) direct causal relation from Q1 to Q2.
Furthermore, Fig.~\ref{fig:Total_causal_effects} also shows that the total causal effect on Q2 was attributed solely to Q1.
Those results suggest that the participants' understanding of the APMV's driving intentions (Q1) positively affects their prediction (Q2) of the APMV's driving behavior.

This result aligned with the process of situation awareness~\citep{endsley1995toward} in the hypothesized model (see Fig.~\ref{fig:model}).
Additionally, \citep{liu2021importance,liehr2024you,kuge2024ehmi} also mentions that a good understanding of the intentions of automation systems can improve the user's ability to predict the behavior of AVs or APMVs.

\subsubsection{Causal Effects on the Trust in APMV}

Moreover, Fig.~\ref{fig:Bootstrap} shows that participants' trust in APMV (Q4) tends to increase if the driving intentions of APMV could be easily understood (Q1) and the driving behavior could be easily predicted (Q2).
This suggests that the pedestrians' trust in an APMV is the situational trust~\citep{hoff2015trust}, primarily affected by their understanding of the APMV's intentions and prediction of its behaviors, rather than by initial trust or trust propensity.

This finding is consistent with the results of \citep{verberne2012trust, petersen2019situational, liu2021importance, avetisyan2022investigating} in the interactions between humans and AV.
In addition, \citet{yang2017evaluating} introduced from another perspective that the higher the transparency of an automation system, the more users tend to trust the system. 
Similarly, \citet{m2021calibrating} used an eHMI to enhance the transparency of an AV system by displaying its status and intentions, which can improve pedestrian trust.
These studies also indirectly suggested that users' situational awareness of the automation system influences their trust in it.

We considered that the above results and conclusions could offer insights into an important topic in human-machine systems: calibrating humans' trust in automation systems.
Specifically, \citet{hollander2019overtrust,kaleefathullah2020external} reported that during interactions between pedestrians and AVs equipped with eHMIs, there is a potential risk that pedestrians develop over-trust in the eHMI and the AV, which could lead to unsafe situations.
Therefore, we consider that, to calibrate the trust of humans (\ie pedestrians in this study) in automation systems, it is necessary to help them calibrate their situation models (see Fig.~\ref{fig:model}).
Furthermore, \citet{endsley2000} suggested that the situation model is supported by the mental model, which provides the necessary prior knowledge for situation awareness.
Meanwhile, \citep{jones2011mental,Al-Diban2012} pointed out that the mental model is a highly organized and dynamic knowledge structure.
It serves as an internal representation of a target system that contains meaningful declarative and procedural knowledge derived from long-term experience and study.
Therefore, in summary, to calibrate the trust of humans (\ie pedestrians in this study) in automation systems (\ie the APMV in this study), it is essential to calibrate their mental model of the automation system.
For example, some studies suggested the use of pre-instruction~\citep{liu2021importance} and educational HMI~\citep{matsuo2024enhancing} to help users develop an accurate mental model.

\subsubsection{Causal Effects on the sense of relief}

From Fig.~\ref{fig:Bootstrap}, we found that pedestrians feel relief (Q5) during interactions with an APMV only when they have established sufficient trust (Q4) in it, as Q4 was identified as the only direct causal effect on Q5, with a reproducibility probability of 100\%.
This finding not only provides empirical support for \citep{Matsubayashi_2023}, which argued that ``anshin'' (corresponding to Q5, sense of relief, in this study) is highly dependent on trust, but it can also help explain the results reported in \citep{liu2021importance}, which showed that higher pedestrian trust in AVs is associated with a greater sense of relief when crossing the road.

Moreover, from the total causal effects on Q5, Fig.~\ref{fig:Total_causal_effects} shows that, in addition to the direct causal effect from the trust in APMV (Q4), understanding the APMV's driving intentions (Q1) and predicting its driving behavior (Q2) also have indirect effects on the sense of relief (Q5).
This result further shows that pedestrians' trust, based on their understanding of the APMV's driving intention, improves their sense of relief during interactions with the APMV.

\subsubsection{Causal Effects on the sense of danger} 

We found that Q1, Q2, and Q4 have negative direct causal effects on Q3, as shown in Fig.~\ref{fig:Bootstrap}.
Similarly, Fig.~\ref{fig:Total_causal_effects} shows that only the above three factors (Q1, Q2, and Q4) have negative total causal effects on Q3.
These findings suggest that participants experience a marked increase in perceived danger when they face challenges in understanding the APMV's driving intentions, encounter difficulties in predicting its driving behavior, and exhibit a lack of trust in the APMV.
This result corresponds to \citep{liu2021importance}, noting that inadequate situational awareness of AVs heightens pedestrians' perceived danger. Moreover, \citet{Clercq2019eHMI} demonstrated that clear AV driving intentions reduce this perceived risk. 
Similarly, \citet{liu2022implicit} found that diminished comprehension of APMV driving intentions raises the conditional probability of perceived danger.
Moreover, \citet{kenesei2022trust} observed that trust in the AV's performance can reduce the user's perceived risk, which aligns with our findings. In contrast, \citet{ZHANG2019207} reported that perceived risk can inversely affect users' initial trust in AVs.
We hypothesize that, in human-machine systems, initial trust is influenced by subjective risk.
Once trust reaches a certain level, it begins to inversely affect subjective risk.
This hypothesis needs to be further validated in future studies.

\subsubsection{Causal Effects on the hesitation in decision-making}

For both the direct and total causal effects, as shown in Fig.~\ref{fig:Bootstrap} and Fig.~\ref{fig:Total_causal_effects}, Q6 was negatively influenced by Q1, Q2, and Q4, while being positively influenced by Q3. This suggests that participants showed increased hesitation in decision-making when they struggled to comprehend the APMV's driving intentions, predict its behavior, had lower trust, and perceived higher danger.

This aligns with \citet{liu2022implicit} regarding pedestrians' subjective evaluations during encounters with AVs. 
However, \citep{liu2022implicit} did not clarify the causal relations between these factors.
Additionally, \citet{yang2024interpreting} found that when pedestrians could better understand the AV's intentions through eHMI, they turned their heads less before crossing the street, which means they had greater confidence and reduced uncertainty in the crossing. 
The confidence in crossing the street observed here can also be interpreted as being negatively correlated with hesitation in decision-making. 
In other words, when pedestrians can more easily understand the AV's intentions, their hesitation in decision-making decreases, which aligns with our finding.

\subsubsection{Causal Effects on Walking behaviors}

Although Q1, Q3, and Q6 were designed to measure participants' subjective experiences during the processes of CIT, CT, and ACT, respectively, and Q5 was designed to measure their sense of relief during CIT and CT, the causal discovery results revealed that Q1, Q3, Q5, and Q6 exhibited a direct causal relation only with CIT (see Figs.~\ref{fig:Bootstrap} and \ref{fig:Total_causal_effects}).
These results suggest that pedestrians' subjective experiences before initiating crossing, such as intention understanding, danger perception, sense of relief, and hesitation, accumulate over time and play a crucial role in their walking behaviors before the crossing, \ie the CIT.
Once the crossing action is initiated, these subjective experiences may no longer significantly influence CT or ACT.

Consequently, the discussion focuses on CIT as the behavior most affected by subjective experiences.
Figs.~\ref{fig:Bootstrap} and \ref{fig:Total_causal_effects} show that CIT was negatively affected by Q1, Q3, Q5, and positively affected by Q6.
This suggests that easier understanding in driving intention and a higher sense of relief allow pedestrians to start crossing the road more quickly, thereby shortening their CIT.
In contrast, hesitation has a positive impact, delaying the initiation of crossing.
The above findings align with the conclusions of \citep{loew2022go,yang2024interpreting}, which suggested that when pedestrians could easily understand the driving intentions of AVs conveyed through eHMIs, their CIT also decreased. 
Furthermore, \citet{lee2022learning, lee2024hello} reported that as pedestrians became more familiar with the information displayed on eHMIs, meaning they were better able to understand these information, then their CIT decreased.



Additionally, there seems to be a perplexing result where the direct causal effect from Q3 to CIT was negative, \ie $-0.29$, shown in Fig.~\ref{fig:Bootstrap}, but the correlation coefficient between Q3 and CIT was positive, \ie $0.19$, shown in Table~\ref{tab:correlation}.
This could be considered because correlation analysis focuses on the relation between two variables, whereas DirectLiNGAM based on SEM involves a multivariate linear regression analysis.
Further evidence lies in the fact that while Q3 had a direct causal effect of $-0.29$ on CIT (a negative influence, meaning decreased perception of danger delays crossing initiation), its total causal effect on CIT is $-0.14$, which was smaller than the direct causal effect. 
This suggests that Q3 has a positive indirect causal effect on CIT through Q6, which partially offsets its negative direct effect in the total causal effect.

\subsection{Effects of eHMI Conditions on Subjective Evaluations and Walking Behaviors based on Causal Discovery Results}

Firstly, an obvious result is that the post-hoc multiple comparisons (see Table~\ref{tab:post-hoc}) did not reveal significant differences between \textit{Early eHMI} and \textit{Sync eHMI} in their effects on the six subjective evaluations and three walking behaviors.
We considered that the limited size of the experimental site (see Fig.~\ref{fig:scenarios}) resulted in a time difference of only about one second between the information cues of \textit{Early eHMI} and \textit{Sync eHMI}.
This minor time difference was likely imperceptible to the participants, as a few participants reported after the experiment that they did not perceive any distinction in the timing of the information cue between \textit{Early eHMI} and \textit{Sync eHMI}.

Secondly, guided by the causal discovery results in Fig.~\ref{fig:Bootstrap}, the results of the post-hoc multiple comparisons in Table~\ref{tab:post-hoc} are further discussed.
Since only under the conditions of \textit{Early eHMI} and \textit{Sync eHMI} was the driving intention of the APMV conveyed to participants via eHMI before the vehicle came to a stop, participants perceived the APMV's driving intentions to be significantly easier to understand (Q1), which directly led to significantly easier to predict the driving behaviors (Q2), in these two conditions compared to \textit{Late eHMI} and \textit{Non eHMI}, respectively.
This suggests that providing anticipated information about future behaviors via eHMI before the APMV stops significantly enhances pedestrians' situation awareness of the interaction, \ie comprehension of the APMV's driving intentions and their ability to predict driving behaviors.

The above result also directly led to both \textit{Early eHMI} and \textit{Sync eHMI} significantly facilitating the building of participants' trust in the APMV (Q4) compared to \textit{Non eHMI} and \textit{Late eHMI}.
Additionally, since stop-related information was displayed on the eHMI after stopping under \textit{Late eHMI} but not under \textit{Non eHMI}, the trust in the APMV was significantly higher in the \textit{Late eHMI} condition than in \textit{Non eHMI}.
This suggests that the driving intention information provided by the eHMIs might contribute to facilitating the building of pedestrians' trust in the APMV, with \textit{Late eHMI} being less effective than \textit{Early eHMI} and \textit{Sync eHMI}, although still better than \textit{Non eHMI}.

This significant improvement in trust also led to participants feeling significantly more relieved (Q5) when interacting with the APMV under eHMI conditions with information cues, \ie \textit{Early eHMI}, \textit{Sync eHMI}, and \textit{Late eHMI}, compared to \textit{Non eHMI}.
This suggests that providing participants with information about the APMV's driving intentions through eHMI during encounters can help participants feel more relief when crossing the road.

Similarly, under the conditions of \textit{Early eHMI}, \textit{Sync eHMI}, and \textit{Late eHMI}, the significant improvements in situational awareness and trust led to a significant reduction in participants' perceived danger (Q3) during the interaction with the APMV, compared to \textit{Non eHMI}. 
Furthermore, considering that in the \textit{Early eHMI} condition, the eHMI information was displayed earlier than the APMV's deceleration, there was no significant difference in perceived danger between \textit{Early eHMI} and \textit{Late eHMI}. 
This suggests that the alignment of explicit and implicit information during the interaction between the APMV and pedestrians impacts the pedestrians' perception of danger.

Under the \textit{Sync eHMI} condition, the synchronized display of the APMV's driving intentions through eHMI enhanced participants' situational awareness, increased their trust in the APMV, and reduced their perceived danger, resulting in a significant reduction in hesitation (Q6) when making decisions, compared to all other conditions.
Additionally, \textit{Early eHMI} significantly reduced participants' hesitation compared to \textit{Late eHMI} and \textit{Non eHMI}, while \textit{Late eHMI} was more effective than \textit{Non eHMI} in reducing hesitation.
Overall, these results suggest that providing eHMI information can reduce participants' hesitation, with \textit{Sync eHMI} being the most effective.

Furthermore, the \textit{Early eHMI} and \textit{Sync eHMI} enhanced pedestrians' situational awareness, builded trust in the APMV, and reduce feelings of danger. 
As a result, they minimized pedestrians' hesitation when making crossing decisions, ultimately shortening their CIT and increasing the efficiency of walking behaviors.

Although Fig.~\ref{fig:Bootstrap} shows that CT and ACT are independent of the subjective evaluation factors, we found that participants' CT and ACT were significantly longer under the \textit{Late eHMI} condition compared to other eHMI conditions, as shown in Table~\ref{tab:post-hoc}.
We considered that only under the \textit{Late eHMI} condition does the APMV continue displaying ``I stopped'' after coming to a stop, until the participant finishes crossing and the vehicle resumes moving. 
This information reassures participants that the APMV will not move, allowing them to cross the road without haste.
Although this finding was not validated by the causal discovery in this study, we will focus on it in future work.

In summary, this study found that it is important to convey information on the APMV's driving intentions to pedestrians through eHMI when there is a change in driving behavior or before such a change occurs. 
Specifically, this information enhances pedestrians' situational awareness and trust in APMVs, reduces perceived risks during interactions, enables quicker decision-making, and shortens the crossing initiation time, thereby improving the efficiency of the interaction.

\subsection{Limitations and Future Work}


First, although the questionnaire (Q1–Q6) was developed based on the hypothesized cognition–decision–behavior process, it has not yet been cross-validated with established scales assessing constructs such as trust or risk perception.
Moreover, this study only examined the causal relations between pedestrians’ subjective experiences and their walking behaviors through a retrospective evaluation, without considering the dynamic changes of these factors during interactions with APMVs. 
In particular, trust in AVs is regarded as a cumulative, dynamic variable~\citep{hu2021trust, walker2023trust}.
Future work will cross-validate the questionnaire and apply other causal discovery methods such as CaPS~\citep{xu2024ordering} and JIT-LiNGAM~\citep{pmlr-v213-fujiwara23a} to discover dynamical and nonlinear causal relations from pedestrians' psychological states to their walking behaviors.

Second, the participants were predominantly young Asian males, which may have influenced subjective experiences and behaviors due to age, cultural, and gender-related factors, limiting generalizability. 
Future works will expand demographic diversity to validate the findings across broader populations and will also investigate how cultural differences may affect the outcomes of the causal relations.

Third, the experimental setting lacks real-world complexity, including only the one-to-one interactions between pedestrians and APMVs, the absence of passengers on the APMV, encounters with APMVs exhibiting fixed driving kinematics and trajectory in each trial, and limited eHMI designs.
Future work will incorporate more realistic scenarios, including varied APMV driving kinematics and eHMI designs, and incorporate APMV's driving behaviors and the variety of messages conveyed by eHMIs as factors in the causal discovery process, aiming to construct a bidirectional causal model that captures the mutual influences between pedestrians and APMVs during interactions.
Then, based on the discovered causal relations, eHMI design guidelines will be developed to enhance pedestrian-APMV interactions.
Moreover, future work will also investigate the effects of APMV's passengers on pedestrian-APMV interactions.

Fourth, walking durations (\ie CIT, CT, ACT) were measured using OpenPose without accounting for perspective distortion or participant height, which may have affected measurement accuracy. 
In future work, improving measurement techniques, such as using ground-based LiDAR, can enhance the metrics' accuracy of walking behaviors.

Finally, although this study preliminarily incorporated pedestrians' feelings of hesitation in decision-making within the causal model, it did not explore the underlying decision-making processes in depth. 
Thus, future work will reference established decision-making models in pedestrian-vehicle interactions (\eg \citep{markkula2023explaining, Tian2025}) to develop a more detailed and comprehensive causal model that captures the progression from pedestrians' psychological states to their walking behaviors.

\section{CONCLUSION}
This study investigated the causal relations from pedestrians' subjective evaluations to their walking behavior during interactions with APMV.
In the experiment, various eHMIs conditions were designed to induce participants to experience different levels of subjective evaluations and generate corresponding walking behaviors.
DirectLiNGAM was used for causal discovery and the results were consistent with the hypothesized model shown in Fig.~\ref{fig:model}. 
Furthermore, the experimental results enriched the detailed causal relations in the hypothetical model, such as how the outcomes of situation awareness led to a sense of danger, trust in APMV, and a sense of relief; how situation awareness, the sense of danger, and trust in APMV contributed to hesitation in decision-making; and how situation awareness, the sense of danger, and hesitation influenced walking behaviors.
Finally, this study found that when the APMV conveyed its driving intentions to pedestrians through eHMI during or before changes in driving behavior, it enhanced subjective evaluations of pedestrians and made their walking behaviors more efficient during interactions.

\section*{ACKNOWLEDGMENTS}
This work was supported by JSPS KAKENHI Grant Numbers 20K19846 and 22H00246, Japan. 

\section*{CRediT Author Statement}
\textbf{Hailong~Liu}: Conceptualization, Investigation, Methodology, Software, Validation, Formal analysis, Project administration, Funding acquisition, Writing - Original Draft.
\textbf{Yang~Li}:  Methodology, Investigation, Writing - review \& editing.
\textbf{Toshihiro~Hiraoka}: Conceptualization, Methodology, Writing - review \& editing.
\textbf{Takahiro~Wada}: Methodology, Writing - review \& editing.

\footnotesize
\bibliographystyle{IEEEtranN}
\bibliography{sample.bib}

\vspace{-9mm}
\begin{IEEEbiography}
[{\includegraphics[width=1in,height=1.25in,clip,keepaspectratio]{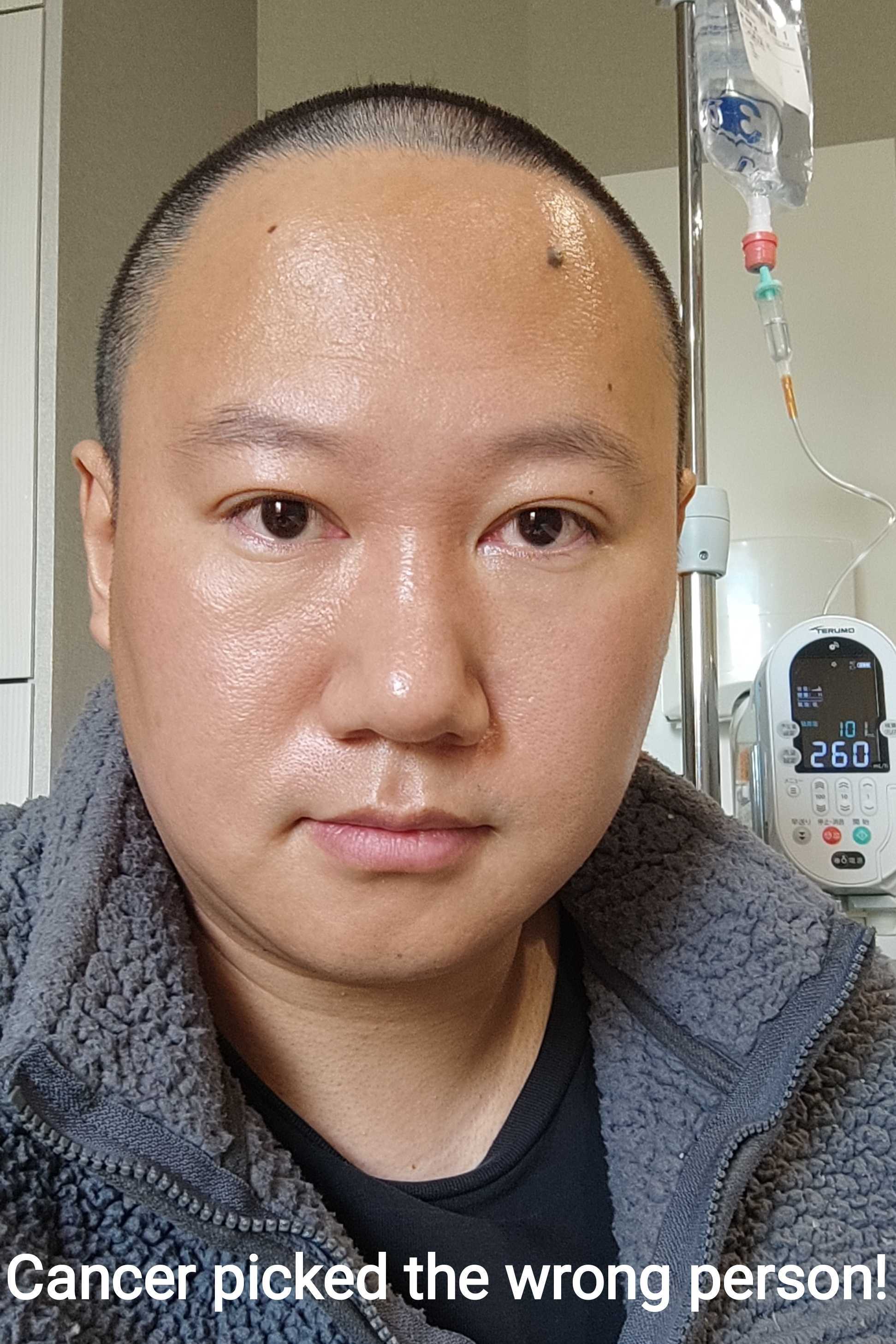}}]
{Hailong~Liu} (S'15--M'19--SM'25) received his M.Eng. and Ph.D. degrees in Engineering from Ritsumeikan University, Japan, in 2015 and 2018, respectively. He was a JSPS Research Fellow for Young Scientists (DC2) (2016--2018), a researcher at Nagoya University (2018--2021).
In Nov. 2021, he joined Nara Institute of Science and Technology (NAIST), Japan, as an Assistant Professor and was promoted to Associate Professor in Feb. 2024.
From Oct. 2025, he becomes a cancer warrior.
His research focus on human factors and machine learning in intelligent transportation systems. He is a Senior Member of IEEE and holds memberships in IEEE ITSS, RAS, SMC. He also serves on the Human Factors in ITS Committee of IEEE ITSS. In addition, he is a member of JSAE, JSAI, and SICE.
\end{IEEEbiography}
\vspace{-9mm}

\begin{IEEEbiography}
[{\includegraphics[width=1in,height=1.25in,clip,keepaspectratio,trim=0 200 0 100]{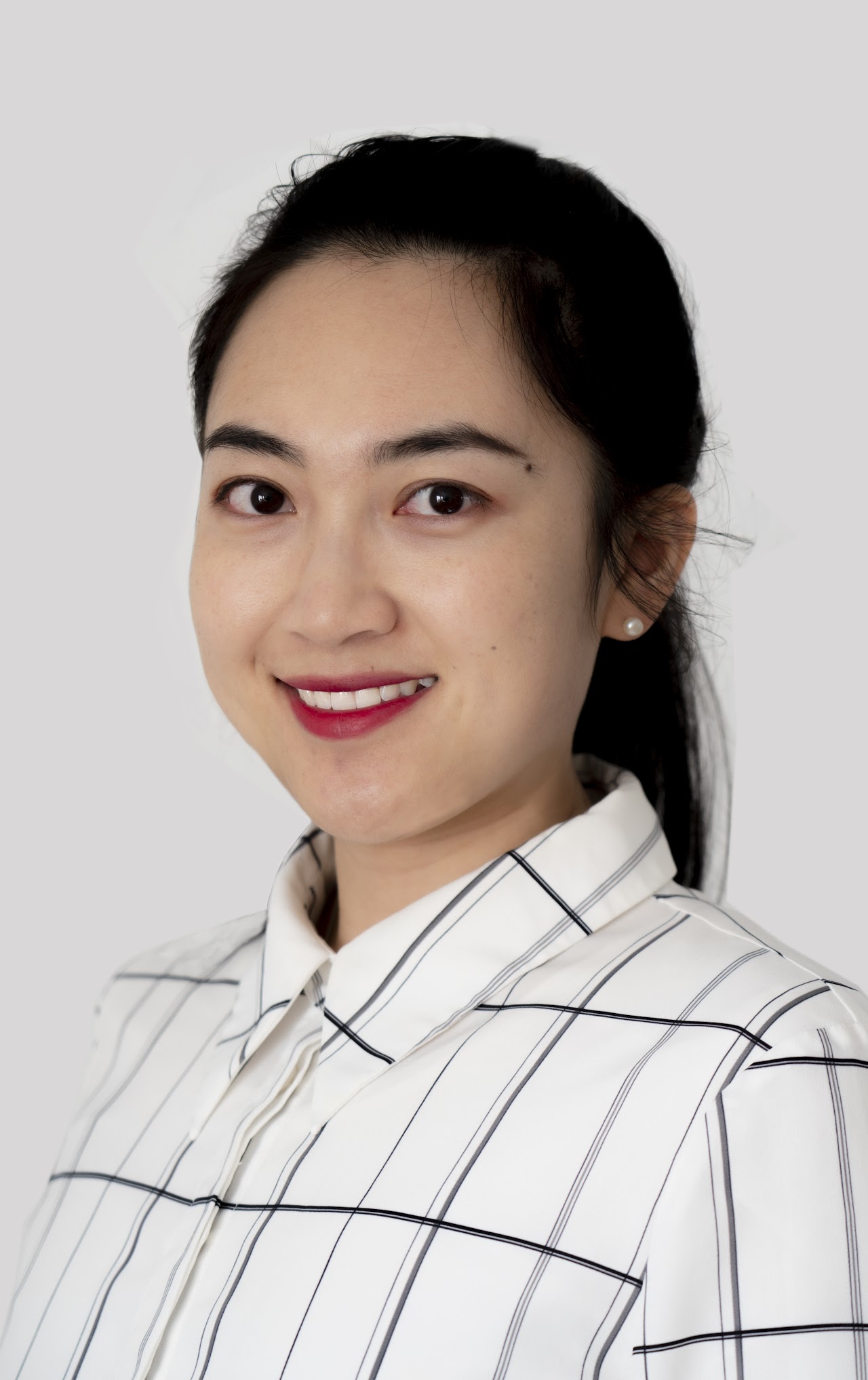}}]
{Yang~Li}
is currently pursuing the Ph.D. degree with
Karlsruhe Institute of Technology (KIT), Germany.
Her research focuses on human-machine interfaces
that facilitate communication between automated
vehicles (AVs) and human traffic participants in
ambiguous road scenarios. She completed a 1-year
visiting study at the University of Leeds and took
part in the Hi-Drive Program in 2022. In 2023,
she also spent one month at the Nara Institute
of Science and Technology (NAIST) as a visiting
research student.
\end{IEEEbiography}
\vspace{-9mm}

\begin{IEEEbiography}
[{\includegraphics[width=1in,height=1.25in,clip,keepaspectratio,trim=20 0 30 20]{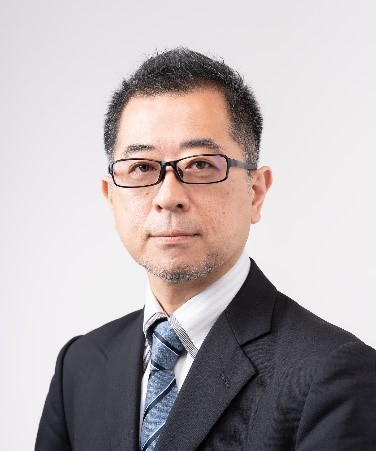}}]
{Toshihiro~Hiraoka} (M'14) received B.E. and M.E. degrees in Precision Engineering in 1994 and 1996, and a Ph.D. in Informatics in 2005, all from Kyoto University, Japan.
He worked at Matsushita Electric Industrial Co., Ltd. (1996–1998), Kyoto University as an Assistant Professor (1998–2017), Nagoya University as a Designated Associate Professor (2017–2019), and the University of Tokyo as a project professor (2019–2022). Since 2022, he has been a Senior Chief Researcher at the Japan Automobile Research Institute. His research interests include human–machine systems, advanced driver-assistance systems, and automated driving systems. He is a member of SICE, HIS, JSAE, JES, IATSS, and IEEE (ITSS).
 \end{IEEEbiography}
\vspace{-9mm}

\begin{IEEEbiography} [{\includegraphics[width=1in,height=1.25in,clip,keepaspectratio]{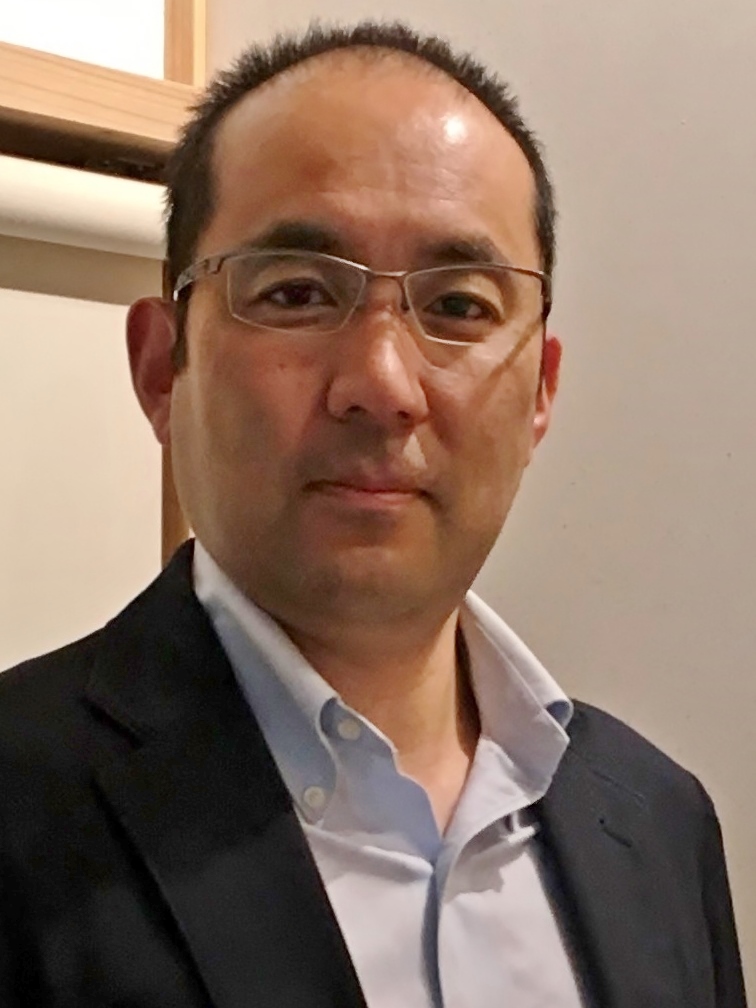}}]
{Takahiro~Wada} 
(M'99) received a B.S. degree in Mechanical Engineering, a M.S. degree in Information Science and Systems Engineering, and a Ph.D. degree in Robotics from Ritsumeikan University, Japan, in 1994, 1996, and 1999, respectively. 
He as a Research Associate worked at Ritsumeikan University (1999-2000).
He worked at Kagawa University as a Research Associate (2000-2003), an Assistant Professor (2003--2007) and Associate Professor (2007--2012).  
He has been a full professor at Ritsumeikan University (2012--2021) and at Nara Institute of Science and Technology (2021--present).
His current research interests include robotics, human machine systems, and motion sickness modeling.
He is a member of  IEEE (RAS, ITSS, SMC), SAE, HFES, SICE, JSAE, RSJ, JSME.
\end{IEEEbiography}

\end{document}